\documentclass[aps,prc,reprint,showpacs,nofootinbib,superscriptaddress]{revtex4-2} %nofootinbib,

\usepackage[utf8]{inputenc}     % Para caracteres UTF-8
\usepackage[T1]{fontenc}        % CodificaciÛn de fuente
\usepackage{amsmath,amssymb}    % Paquetes matem·ticos
\usepackage{graphicx}           % InclusiÛn de im·genes
\usepackage{dcolumn}            % AlineaciÛn decimal en tablas
\usepackage{bm}                 % SÌmbolos matem·ticos en negrita
\usepackage{hyperref}           % Enlaces y referencias con hipervÌnculos
\usepackage{color}              % Texto en color (opcional)

\def\bvl{\hbox to0.4pt{\leaders\hrule height6pt\hfill}}
\def\swb{\hbox to6pt{\hfill}}
\def\bhl{\hbox to 6.8pt{\leaders\hrule\hfill}}

\def\xbox{\hbox{$\!$\raise-1pt\hbox{
            \vbox{\offinterlineskip\bhl\hbox{\bvl\swb\bvl}\bhl}}}}

\def\no{\noindent}

\def\vr{\boldsymbol{\mathrm{r}}}
\def\valpha{\boldsymbol{\mathrm{\alpha}}}
\def\vsigma{\boldsymbol{\mathrm{\alpha}}}
\def\vnabla{\boldsymbol{\mathrm{\nabla}}}
\def\vl{\boldsymbol{\mathrm{l}}}
\def\vs{\boldsymbol{\mathrm{s}}}
\def\vL{\boldsymbol{\mathrm{L}}}
\def\vS{\boldsymbol{\mathrm{S}}}
\def\vrho{\boldsymbol{\mathrm{\rho}}}

\def\e{\epsilon}
\def\no{\noindent}

\def\be{\begin{equation}}
\def\ee{\end{equation}}
\def\ba{\begin{array}} 
\def\ea{\end{array}}
\def\bea{\begin{eqnarray}} 
\def\eea{\end{eqnarray}}

\begin{document}

\title[Article Title]{Charge radii of Sn isotopes in the relativistic mean field approximation}

\author{S. Marcos}

\email{Contact author: marcoss@unican.es}

\affiliation{Departamento de F\'\i sica Moderna, Universidad de Cantabria, E-39005 Santander, Spain.}

\author{N. Sandulescu}

\email{Contact author: sandulescu@thery.nipne.ro}

\affiliation{{National Institute of Physics and Nuclear Engineering}, RO-077125 M\u{a}gurele, Romania.}

\author{R. Niembro}

\email{niembror@unican.es}

\affiliation{Departamento de F\'\i sica Moderna, Universidad de Cantabria, E-39005 Santander, Spain.}

\date{\today}

\begin{abstract}
The kink observed in the nuclear charge radius of Sn isotopes around neutron number $N = 82$ is investigated within the relativistic mean-field (RMF) framework using the NL3$^*$ parameter set. 
It is shown that the small components of the Dirac spinors for the neutron single-particle states near the Fermi level play a crucial role in forming the kink through their contribution to the proton central potential. 
In particular, the significant differences between the radial parts of the small components of spin--orbit partner states make neutrons with $j = l - 1/2$ more efficient in increasing the nuclear charge radius than those with $j = l + 1/2$. 
However, the effect induced by the small components alone does not fully account for the magnitude of the kink observed in Sn isotopes. 
\end{abstract}

%\keywords{Relativistic models, nuclear charge radius, kink effect}
%\pacs{21.10.Ft, 27.80.+w}

\maketitle

\section{Introduction}

The nuclear root-mean-square (rms) charge radius ($R_c$) is a fundamental macroscopic property of atomic nuclei.
Traditionally, rms charge radii have been measured using electron scattering.
However, recent advances in laser spectroscopy enable highly precise measurements of subtle differences in $R_c$ between neighboring isotopes, differences that lie beyond the resolution of electron scattering and the data obtained from muonic atom spectra \cite{Ang13, Cam16}.
This development highlights the growing relevance of variations in $R_c$ between neighboring nuclei along isotopic chains.

It is well established that certain isotopic chains exhibit a pronounced change in the behavior of $R_c$ when the neutron number crosses a specific threshold. 
This effect typically occurs when the neutron number ($N$) passes through a magic shell closure, particularly at $N = 28, 50, 82,$ and $126$ \cite{Ang13, Gar16, Gor19, Gar20, Sha93, Yue24}. 
A prominent example of this trend is found in the Pb isotopes \cite{Sha93}, where a marked change in $R_c$ appears once $N$ exceeds 126. 
Other isotopic chains in which $R_c$ exhibit a similar behavior, particularly at $N = 126$, are reported in Ref.~\cite{Yue24}. 
Further examples showing comparable $R_c$ trends for $N > N_{\text{magic}}$ and significant deviations for $N < N_{\text{magic}}$ are discussed in Ref.~\cite{Gar20}. 
The pronounced change in $R_c$ within an isotopic chain, as discussed above, is commonly referred to as the ``kink'' or ``kink effect'' (KE).

The kink in Pb isotopes is accurately reproduced by the relativistic mean-field (RMF) approximation  (RMFA) \cite{Sha93, Mar01, Nie12, Agb14, Per21, Nai23}, whereas nonrelativistic Skyrme-Hartree-Fock (SHF) models with standard parametrizations fail to capture it \cite{Taj93}.
However, both RMF and SHF approaches have difficulties to reproduce the kink observed in the Sn isotopic
chain at $N = 82$ \cite{Nak19, Nai23}, although the former provides a better description than the latter. 
A more satisfactory description is achieved within the relativistic Hartree-Fock approach with density-dependent couplings \cite{Wan21, Nai23}. In this case, the inclusion of the pion-nucleon tensor interaction reduces the energy spacing between the neutron $2f_{7/2}$ and $1h_{9/2}$ orbitals, leading to an improved reproduction of the observed kink.

The different predictions of relativistic and nonrelativistic mean-field models regarding the kink have been examined in several studies.
Reference \cite{Rei95} argues that the failure of standard Skyrme interactions to reproduce the kink arises from an overly strong spin--orbit interaction, which stems from the isospin dependence of the Skyrme functionals.
To address this problem, modified Skyrme functionals have been proposed \cite{Rei95, Sha95}, adjusting the relationship between the isovector ($W'_0$) and isoscalar ($W_0$) components of the spin--orbit interaction in order match closely the one employed in the RMFA. However, as shown in Refs. \cite{God13, Nai23}, a reasonable reproduction of the kink can also be achieved while retaining the standard condition $W'_0 = W_0$, provided that the value of $W_0$ is chosen to ensure a high occupancy of the neutron $1i_{11/2}$ orbital.

An alternative nonrelativistic mean-field model proposed by Fayans {\it et al.} \cite{Fay94, Fay96, Fay00, Rei17, Gor19, Gus25} introduces density gradient terms in both the surface and pairing part of the energy functional. After appropriate parameter adjustments, this model 
predicts a reasonable kink both in Pb and Sn isotopes, although it tends to exaggerate the curvature of the function $Rc(^A\text {Sn})$ in tin isotopes for $A<132$ \cite{Gor19, Gus25}. 
However, the microscopic justification for these gradient-dependent terms has been questioned \cite{Per21}.

Studies of the KE using nonrelativistic finite-range density functionals are scarce. 
Based on the nonrelativistic Michigan three-range Yukawa (M3Y)-~type interaction, 
Nakada {\it et al.} introduced a density-dependent component into the spin--orbit interaction \cite{Nak15, Nak15R, Nak19}, 
leading to the parametrization M3Y-P6a \cite{Nak15}. 
This modification, inspired by chiral effective field theory and associated with three-nucleon forces, 
enhances the magnitude of the kink in $R_c$ of tin isotopes, 
bringing it closer to experimental values, 
while the kink in Pb and Hg isotopes remains too small \cite{Nak20, Day21a}.     
%Yukawa-type interactions with density-dependent spin--orbit terms have also been successfully applied at the Hartree--Fock--Bogoliubov level to investigate $R_c(A)$ in Ca, Ni, Sn, and Pb isotopes~\cite{Nak15R, Nak15, Nak19}.
The structural evolution of $R_c$ in Sr, Zr, and Mo isotopes within the mean-field approximation based on the D1S-Gogny interaction was analyzed in Ref.~\cite{Rod10}. 
The steep behavior observed in the $R_c(A)$ function for Sr and Zr isotopes was associated with nuclear shape transitions. 
Using the same interaction, a detailed study of $R_c(A)$ for Ca isotopes was performed in Ref.~\cite{Gar16}. 

Several research groups have recently reported significant progress in understanding the mechanisms underlying the kink effect. 
For instance, Ref. \cite{Hor22} discusses how core swelling induced by pairing interactions can globally reduce the radii of valence neutron orbitals.
Reference~\cite{Nai23} highlights the relevance of two factors: the tensor interaction in relativistic models, which affects the spin--orbit potential and thereby the occupancy of single-particle neutron states, and the symmetry energy (see also Ref.~\cite{Mar24a}), which depends on the neutron-proton interaction.
%However, as noted in Ref. \cite{Mar23}, the overall impact of tensor interactions remains small for a fixed neutron configuration.
References~\cite{Day21a, Day21b} emphasize that, in addition to the symmetry energy, it is necessary to take into account that the particle-vibration coupling in odd-$A$ nuclei modifies their single-particle (sp) energy spectra, which, consequently, can differ significantly from those of neighboring even-$A$ nuclei.

The crucial role of the neutron $1i_{11/2}$ orbital in the kink behavior observed in Pb isotopes has been highlight in several Refs. \cite{Mar01, God13, Per21, Per23, Nai23}.
In Refs.~\cite{God13, Per23}, this role is attributed
to its strong spatial overlap with nodeless proton orbitals that predominantly drive the KE \cite{Nie12, God13}. This interpretation is based on the assumption that spatial overlap directly correlates with the strength of the neutron-proton interaction in these orbitals, an assumption which was discussed in detail in Ref. \cite{Mar24a}. 
There, a more reasonable interaction (from the nonrelativistic point of view) between neutrons and protons, 
based in the overlaps of their corresponding densities, were analyzed. However, it was not possible to establish a clear correlation between these overlaps and the magnitude of the kink in the RMFA. 

Recently, it has been shown that within the RMFA, the small components of the 
sp Dirac spinors play a crucial role in predicting the observed KE
in the lead isotopic chain \cite{Mar24b}. Furthermore, it is explained why valence neutrons in the $2g_{9/2}$ orbital are less effective in generating the kink than those in the $1i_{11/2}$ orbital. 

A comprehensive overview of the efforts undertaken prior to 2021 to understand the charge radius behavior across various isotopic chains is provided in Ref. \cite{Per21}.

In this paper, we investigate the mechanism underlying the kink formation in the Sn isotopic chain within the RMFA. 
Our analysis focuses on how valence neutrons influence the charge radius through their impact on the effective single-particle central potential. 
In particular, we analyze how the valence neutron orbitals $1h_{9/2}$ and $2f_{7/2}$ contribute to the 
emergency of the kink, distinguishing explicitly between the roles of their large and small components.

The method of analysis was previously applied to lead isotopes in Ref.~\cite{Mar24b}. 
In the present work, we extend and adapt this approach to the study of tin isotopes, 
providing additional methodological details and implementing an improved treatment of
pairing correlations---particularly in the vicinity of closed shell---through 
particle-number-projected BCS. 
Although the procedures adopted for the Pb and Sn isotopic chains are formally similar, the results obtained
for Sn cannot be directly inferred from the earlier Pb study, since the underlying 
single-particle structure---particularly in the kink region---differs significantly
between the two cases.

The aim of the present study is to assess whether the conclusions established for the lead isotopes
also hold for tin, and to clarify whether the origin of the kink observed in these isotopic chains
within the RMFA constitutes an intrinsic feature of the relativistic framework.

The paper is organized as follows.
Sec.~II presents a brief overview of the relativistic model and the Schr\"odinger-like equation equivalent to the Dirac equation used in our calculations. In Sec.~III we discuss the results for the charge radii of Sn isotopes. 
Finaly, in Sec.~IV we present the conclusions.

\no
\section{The Relativistic Model}

In this work, we describe the charge radii within the framework of the RMFA~\cite{Bog77, Bou84, Ruf88, Rin96}.
The calculations are performed with the NL3$^*$ 
parameter set \cite{Lal09} and by using  the no-sea approximation, which means that the processes involving the creation or annihilation of nucleon-antinucleon pairs by the nucleon field are excluded \cite{Bou87, Bro78, Sav04}. Within this approach, the time-independent Dirac spinor $\psi_a(\vr)$, representing a nucleon of rest mass $M$ in the state $a$, satisfies the time-independent Dirac equation \cite{Bou84, Rin96}.
\be
\{-i\hbar \valpha\cdot\vnabla + \beta[M+S(\vr)]
+V(\vr)\}\psi_a(\vr)=E_a\psi_a(\vr),
\label{Dir}
\ee
where $E_a = M + \e_a$ and $-\e_a$ is the binding energy corresponding to the state $\psi_a(\vr)$.
The scalar $S(\vr)$ and vector $V(\vr)$ sp potentials, respectively, arise from the exchange of the effective scalar ($\sigma$) and vector ($\omega$, $\rho$, $\gamma$) bosons between nucleons.
They are defined as:
\bea
S(\vr)=g_\sigma\sigma(\vr),
%\nonumber
\label{eqS}
\eea
\bea
V(\vr)=
g_\omega \omega_0(\vr)+\tau_3g_\rho\rho_{0,3}(\vr)+\frac{1+\tau_3}{2}V_{\rm C}(\vr),
\label{eqV}
\eea
\no
where $\sigma(\vr)$ is the static scalar mean field associated with the effective $\sigma$-meson.
The functions $\omega_0(\vr)$ and $\rho_{0,3}(\vr)$ are the time components of the static mean fields $\omega_\mu$ and $\vrho_\mu$, associated with the $\omega$- and $\rho$-mesons, respectively.  
$V_{\rm C}$ denotes the Coulomb potential. The arrow on the $\vrho_\mu$ field indicates that it is a vector in isospin space.  
The constants $g_\sigma$, $g_\omega$, and $g_\rho$ represent the coupling strengths between the nucleon field and the $\sigma$, $\omega$, and $\rho$ meson fields, respectively.  
The isospin projection $\tau_3$ takes the value $+1$ for protons and $-1$ for neutrons.

The effective density Lagrangian corresponding to the NL3$^*$ set includes a potential energy term with cubic and quartic self-interactions in the scalar field \cite{Bog77, Lal09}: $\frac{1}{3}b\sigma^3(\vr) + \frac{1}{4}c\sigma^4(\vr)$, where $b$ and $c$ are parameters fitted within the model. The static meson field equations can be written as \cite{Bou84, Rin96, Ruf88}:
\be
[-{\nabla}^2+{{m_\sigma^*}^2(\vr)}]\sigma(\vr)=-g_\sigma\rho_S(\vr),
\label{eqcs}
\ee
\be
[-{\nabla}^2+m_\omega^2]\omega_0(\vr)=g_\omega\rho_B(\vr),
\label{eqcw}
\ee
\be
[-{\nabla}^2+m_\rho^2]\rho_{0,3}(\vr)=g_\rho\rho_3(\vr),
\label{eqcr}
\ee
\no
where 
\be
{{m_\sigma^*}^2(\vr)}=m_\sigma^2+b\sigma(\vr)+c\sigma^2(\vr),
\ee
\no
and $m_\sigma$, $m_\omega$, and $m_\rho$ are the masses of the $\sigma$-, \mbox{$\omega$-,} and $\rho$-mesons, respectively. 
The functions $\rho_S(\vr)$, $\rho_B(\vr)$, and $\rho_3(\vr)$ denote the scalar, nucleon or baryon, and isovector densities, respectively (see below).

For spherical nuclei, considered in this work, all densities, fields and potentials  have spherical symmetry.  

In the standard notation, $\psi_a(\vr)$ is expressed as
\begin{eqnarray}
\hspace*{-8 pt}\psi_a({\vr})
=\begin{bmatrix}\phi_a(\vr)\\
\zeta_a(\vr)\end{bmatrix}
=\frac{1}{r}\begin{bmatrix}iG_a(r)\cr                
F_a(r){\vsigma}\cdot{\mbox{\boldmath ${\hat{\rm r}}$}}\end{bmatrix}{\it y}^{m_a}_{j_al_a}({\mbox{\boldmath ${\hat{\rm r}}$}})\chi^a_\frac{1}{2},
\label{eqpsi}
%label{eq2}
\end{eqnarray}
\no
where $\frac{G_a(r)}{r}$ and $\frac{F_a(r)}{r}$ represent the radial 
functions of the large and small components,
${\it y}^{m_a}_{j_al_a}({\mbox{\boldmath${\hat{\rm r}}$}})$ denotes the normalized spin-angular wavefunction,
 and $\chi^a_\frac{1}{2}$ is the nucleon isospinor \cite{Bou87, Sak67}. 

The neutron $\rho_{\rm n}(r)$ and proton $\rho_{\rm p}(r)$ densities, and the scalar $\rho_S(r)$ density can be written as  
\be
\rho_{\rm n,\rm p}(r)=\sum_{a\in occ} \frac{2j_a+1}{4\pi}\frac{G_a^2(r)+F_a^2(r)}{r^2},
\label{eqdn}
\ee
\no
where $a$ represents an occupied neutron state for $\rho_{\rm n}(r)$ and an occupied proton state for $\rho_{\rm p}(r)$;
\be
\rho_S(r)=\sum_{a\in occ} \frac{2j_a+1}{4\pi} \frac{G_a^2(r)-F_a^2(r)}{r^2},
\label{eqdS}
\ee
\no
where $a$ refers to any occupied neutron or proton state.
The total baryon density $\rho_B(r)$ and the isovector density $\rho_3(r)$ can be written in terms of $\rho_{\rm n}(r)$ and $\rho_p(r)$ as
\be
\rho_B(r)=\rho_{\rm n}(r)+\rho_{\rm p}(r),
\label{eqdB}
\ee
\be
\rho_3(r)=\rho_{\rm p}(r)-\rho_{\rm n}(r).
\label{eqd3}
\ee

In the case of open-shell nuclei, we also take into account the contribution of pairing
correlations~\cite{Rin80}. They are described with the zero-range pairing interaction
\be
V_P (\vr_1,\vr_2) = - V_0 \delta (\vr_1-\vr_2)
\label{eqVP}
\ee
For the pairing strength we have taken the value $V_0$ = 110 MeV fm$^3$.
With this value, we get for the Sn isotopes an average BCS pairing gap of about 1.5 MeV  around 
the middle of the neutron major shell. 
The effect of the pairing on the mean field is taken into account through the densities. 
More precisely, in the densities given by Eq.~(\ref{eqdn}) and Eq.~(\ref{eqdB}) each term is
multiplied by the occupation probability $v^2_a$ corresponding to the single-particle state $a$. 
The RMF+BCS calculations are done iteratively, at each step updating the occupation probabilities and the densities.

\subsection{The Schr\"odinger-like equation equivalent to the Dirac equation}

To facilitate the discussion of our results, we construct a Schr\"odinger-like equation equivalent to the Dirac equation.
This can be achieved from Eq. (\ref{Dir}) by expressing the small component $\zeta_a(\vr)$ of the Dirac spinors in terms of the large component $\phi_a(\vr)$, and applying the transformation \cite{Ned88}
\be
\phi(\vr)=B(r)^{1/2}\tilde \phi(\vr), 
\label{phi}
\ee
\no
where the subscript $a$ is removed for simplicity, and
\be
B(r)\equiv 2M + \e + S(r) - V(r).
\label{B}
\ee
\no
Then, in terms of $\tilde \phi (\vr)$, the Schr\"odinger-like equation can be written as 
\be
[-\frac{\hbar^2}{2M}\nabla^2+V_{\rm cent}(r,\e)+V_{\rm SO}(\vr,\e)]\tilde \phi (\vr)=
\e\tilde \phi (\vr),
\label{Sch}
\ee
\no
where the central ($V_{\rm  cent}(r,\e)$) and the spin--orbit 
($V_{\rm SO}(\vr,\e)$) potentials are energy dependent \cite{Mar01,Ned88}.  
They read as follows:
\be
V_{\rm cent}(r,\e)= V_{\rm cent^*}(r)+V_{\rm cent}^{\rm rel}(r,\e),
\label{Vcent}
\ee
\be
V_{\rm cent^*}(r) = S(r)+V(r),
\label{Vcent*}
\ee
\bea
\label{Vcentrel}
V_{\rm cent}^{\rm rel}(r,\e) = & \dfrac{S^2-V^2+\e^2}{2M}+\e\dfrac{V}{M} \nonumber \\
+ & \dfrac{\hbar^2}{2M}\left[\dfrac{1}{4}W^2+\dfrac{1}{r}W+\dfrac{1}{2}W'\right],
%\label{Vcentrel} aqui no funciona
\eea
\be
W(r,\e)=-\frac{S'-V'}{2M+\e+S-V};\nonumber
\ee
\be
V_{\rm SO}(\vr,\e)=\frac{{\hbar}^2}{2M}\frac{2}{r}W(r,\e)\vl \cdot \vs\equiv V^r_{\rm SO}(r,\e)\vl \cdot \vs.
\label{VSO}
\ee
\no
Here, $\vl=\vL/\hbar$ and $\vs=\vS/\hbar$, with $\vL$ and $\vS$ being the orbital angular momentum and spin operators of a nucleon, respectively. 

By resolving Eq. (\ref{Sch}), the large component $G(r)$ of a Dirac spinor can be obtained from Eq. (\ref{phi}).
The corresponding small component $F(r)$ can then be derived from $G(r)$ using the relation \cite{Sak67,Bro78}
\be
F(r)=\frac{\hbar[G'(r)+(k/r)G(r)]}{B(r)},
\label{F} 
\ee         
where $k=j+1/2$ for the states with $j=j_-\equiv l-1/2$ and $k=-(j+1/2)$ for states with $j=j_+\equiv l+1/2$. 
The normalization condition for $G(r)$ and $F(r)$ reads
\be
\int[G^2(r)+F^2(r)]dr=1.
\label{nor}
\ee
The functions $G(r)$ and $F(r)$ obtained in the way described above satisfy the Dirac Eq. (\ref{Dir}).
However, if  $F(r)$ is neglected in Eq. (\ref{nor}), Eqs. (\ref{Sch}-\ref{VSO}, \ref{nor}) represent a nonrelativistic approximation to Eq. (\ref{Dir}). 

\subsubsection{Modified spin--orbit interaction}

Various studies indicate that a density-dependent spin--orbit (SO) interaction can significantly affect 
charge radii~\cite{Nak19,Nak20,Nai23}. To investigate if this is also the case in the RMFA, 
in addition to the standard SO interaction $V_{\rm SO}(\vr,\e)$ given by Eq.~(\ref{VSO}), 
we also consider the following density-dependent SO interaction
\begin{align}
{\tilde V}_{\rm SO}(\vr,\e) &= 1.05 
\left( \frac{\rho_B(r)}{\rho_{B0}} \right)^{1/4} V_{\rm SO}(\vr,\e) \nonumber \\ 
&\equiv {\tilde V}^r_{\rm SO}(r,\e)\,\vl\cdot\vs,
\label{VSOM}
\end{align}
where $\rho_B(r)$ is the baryon density and 
$\rho_{B0} = 0.15~\mathrm{fm}^{-3}$ is the nuclear matter saturation density 
corresponding to the NL3$^*$ functional.

\begin{figure}[ht]
%\vskip 10 cm
%\leftskip -10 true cm
\begin{center}
%\centering
\includegraphics[width=5.0 cm,angle=-90.0]{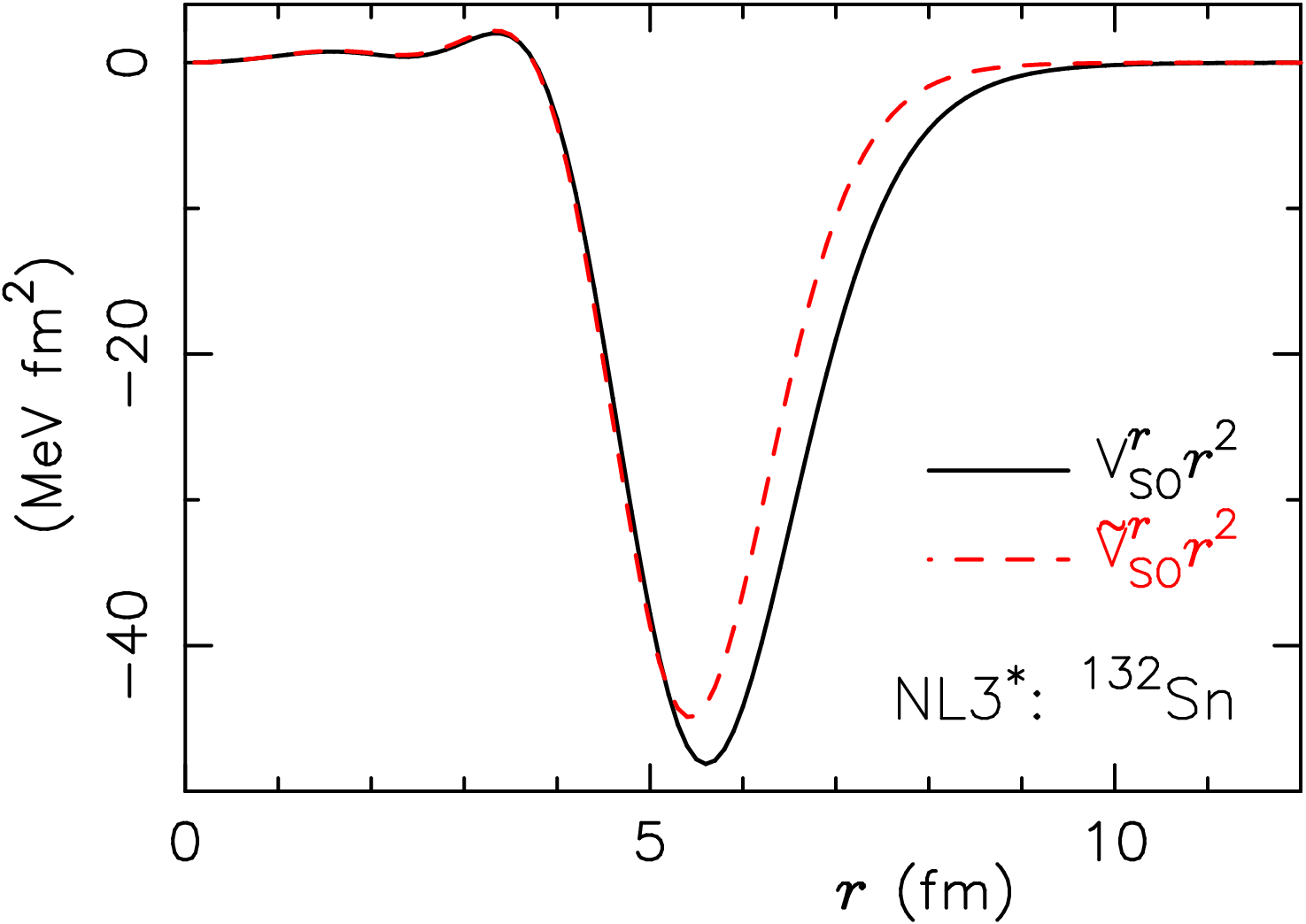} 
\caption{\small Quantities $V^r_{\rm SO}(r,\e)\times r^2$ (Eq.~(\ref{VSO})) and ${\tilde V}^r_{\rm SO}(r,\e)\times r^2$ (Eq.~(\ref{VSOM})) for $^{132}$Sn using the NL3$^*$ parameter set. The energy ($\e$) dependence is negligible.}
\label{VSOVSOM}
\end{center}
\end{figure}

Figure~\ref{VSOVSOM} shows $V^r_{\rm SO}(r,\e)$ and $\tilde{V}^r_{\rm SO}(r,\e)$, each multiplied by $r^2$, to better illustrate their differences in the nuclear surface. 
It is observed that the density-dependent term slightly enhances the standard SO interaction
in the nuclear interior and reduces it near the surface\footnote{The effects of this modification on the SO interaction are correlated with those arising from the introduction of a density dependence in the pion pseudovector coupling constant, $f_\pi$, as done in Refs.~\cite{Lon06, Wan21}. 
This density dependence is chosen to reduce the strength of the nucleon--nucleon tensor interaction in the nuclear interior relative to the surface, which, in turn, indirectly enhances the SO interaction in the interior with respect to that at the surface.
}. It can be also seen that
the most significative differences appear for $r > 5$. It is worth mentioning that the 
transition from the the SO $V_{\rm SO}(\vr,\e)$ to $\tilde{V}_{\rm SO}(\vr,\e)$ is analogous to the transition from Skyrme~M$^*$ to SkI4 reported in Ref.~\cite{Rei95}. 

%Fig. 1

\no
\section{Results}

In this section, we analyze the evolution of the charge radius of Sn isotopes, calculated within the RMFA, as a function of the mass number $A$. This evolution depends on the neutron sp states near the Fermi level and on their 
occupation probabilities.

This section is organized as follows. We begin by presenting the neutron single-particle energy spectrum. We then analyze the evolution of the charge radii when pairing correlations are included within the BCS and projected-BCS (PBCS) frameworks. To clarify the underlying mechanisms, we examine the contribution of the valence neutrons both to the total charge radius and to the radii of selected proton orbitals, considering several representative configurations.
Finally, we assess the respective roles of the large and small components of the valence-neutron Dirac spinors in shaping the charge radius. These effects are interpreted in terms of the corresponding modifications induced in the proton central potential.

\subsection{Single-particle spectrum}

\begin{figure}[ht]
%\leftskip -10 true cm
\begin{center}
%\centering
\includegraphics[width=8. cm,angle=-90.0]{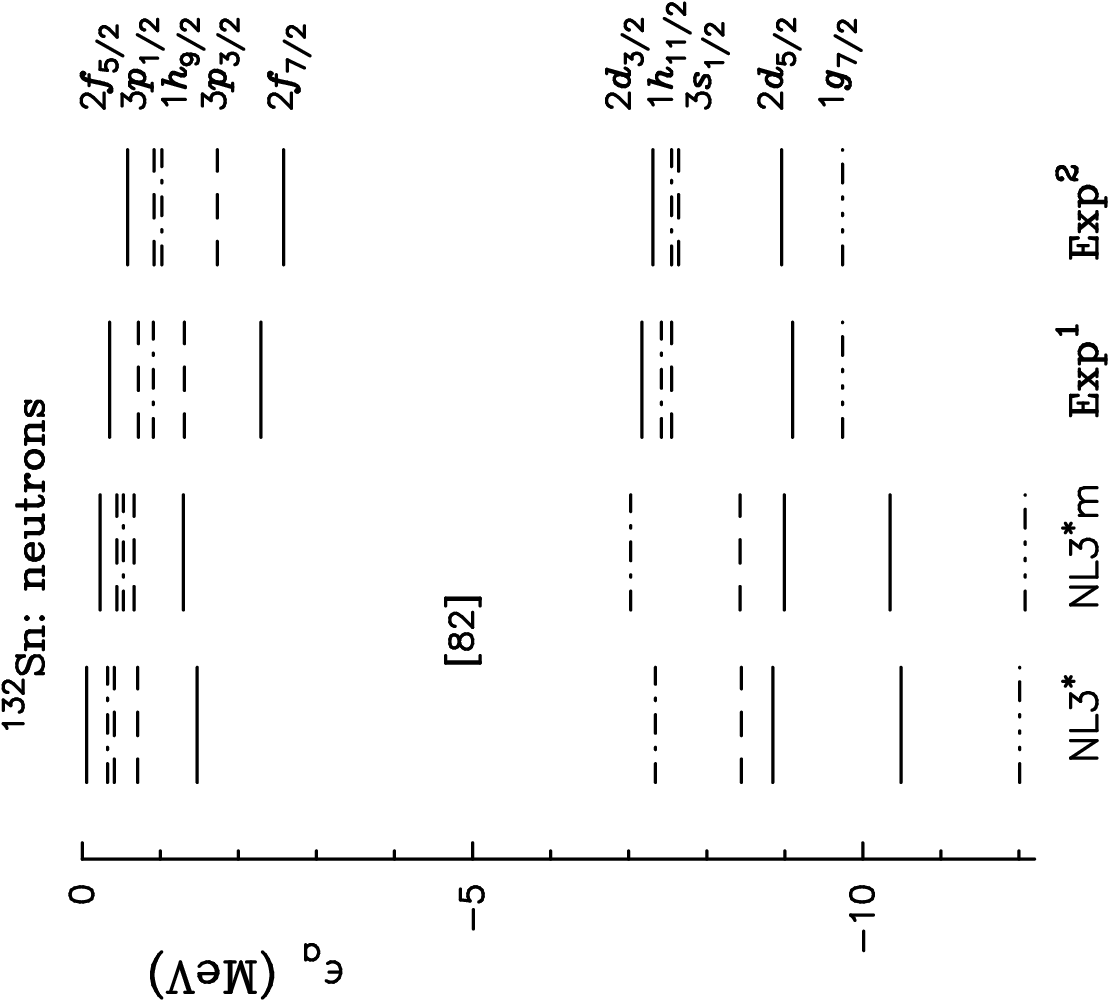}
\caption{\small 
Neutron sp energies of $^{132}$Sn calculated within the RMFA using the NL3$^*$ parameter set. 
The columns labeled by NL3$^*$ (NL3$^*$m) show the results obtained with the standard (modified)
spin--orbit interaction (see the text). The experimental values indicated by
Exp$^1$ and Exp$^2$ are from Refs.~\cite{Oro96} and \cite{Isa07}, respectively.
}
\label{ea}
\end{center}
\end{figure}

Figure~\ref{ea} shows the neutron sp energies corresponding to $^{132}$Sn. 
It includes the states from the major shell $N = 50$ -- $82$ as well as the relevant states above $N = 82$. 
All Sn nuclei discussed in this paper are treated as spherical in our calculations, which appears to be a reasonable assumption (see Ref.~\cite{Lal99}).
%Fig. 2

It can be noticed that the calculated and the experimental sp energies around the Fermi levels are rather different.  
This discrepancy is common to both relativistic and nonrelativistic models and has been attributed, partially, 
to the small value of the energy-dependent effective mass associated with the optical potential \cite{Mah85}. 
In addition, and more importantly, the theoretical sp spectrum is expected to change significantly due to the particle-vibration coupling \cite{Ber80}, which is not considered in this study.

From Fig.~1 it can be seen that the main effect of the modified SO potential given by Eq.~(\ref{VSOM}) 
is a slight reduction in the splittings of the neutron SO doublets above the $N=82$ shell gap, 
leading to an inversion of the $1h_{9/2}$ and $3p_{1/2}$ levels, which lie very close in energy.

\subsection{\it Nuclear charge radii}

\begin{figure}[ht]
%\vskip 10 cm
%\leftskip -10 true cm
\begin{center}
\includegraphics[width=8. cm,angle=0.0]{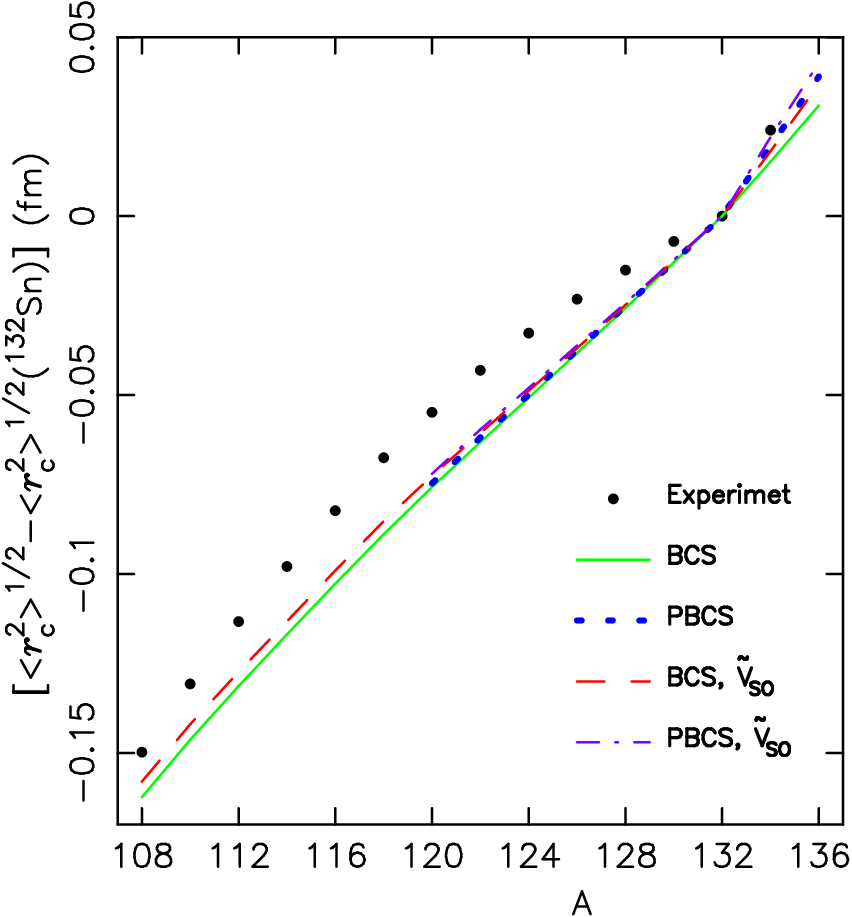}
\caption{\small
Differences of rms charge radii, $\langle r_c^2 \rangle^{1/2} - \langle r_c^2 \rangle^{1/2}(^{132}\text{Sn})$, 
for $^{A}$Sn isotopes calculated within the RMFA using the NL3$^*$ parameter set. 
Curves labeled ``BCS'' (``BCS, $\tilde V_{\rm SO}$'') and ``PBCS'' (``PBCS, $\tilde V_{\rm SO}$'') correspond to results obtained with the standard (modified) spin--orbit interaction, with pairing treated within the BCS and PBCS approaches, respectively. Experimental data are taken from Ref.~\cite{Gor19}.}
\label{difRc}
\end{center}
\end{figure}

The RMF results for the root-mean-square (rms) charge radii ($R_c \equiv {\langle r_c^2 \rangle}^{1/2}$),
relative to the nucleus $^{132}$Sn, are shown in Fig.~\ref{difRc}. The theoretical radii are obtained using the pairing force of Eq.~\ref{eqVP}, treated within the BCS approximation \cite{Rin80}.

For nuclei in the vicinity of $^{132}$Sn, the number of neutrons that effectively participate in pairing correlations is small. Consequently, the BCS solution, which does not conserve particle 
number, is characterized by sizable particle-number fluctuations. To assess how this affects the kink behavior, for the tin nuclei with $A \geq 120$, we have also performed projected-BCS (PBCS) calculations \cite{Die64}, in which particle number is conserved exactly.
Previous studies have shown that, when applied near closed shells, the PBCS approach yields pairing correlation energies and occupation probabilities in very good agreement with the exact solution (see, e.g., \cite{San08}). 

The BCS and PBCS calculations were performed with the single-particle states from the major neutron shells.
Specifically, for  nuclei lighter (heavier) than $^{132}$Sn, we considered the states below (above) the N=82 shell closure, as indicated in Fig. 2.

%Fig. 3

\begin{figure}[ht]
%\vskip 10 cm
%\leftskip -10 true cm
\begin{center}
%\centering
\includegraphics[width=5. cm,angle=-90.0]{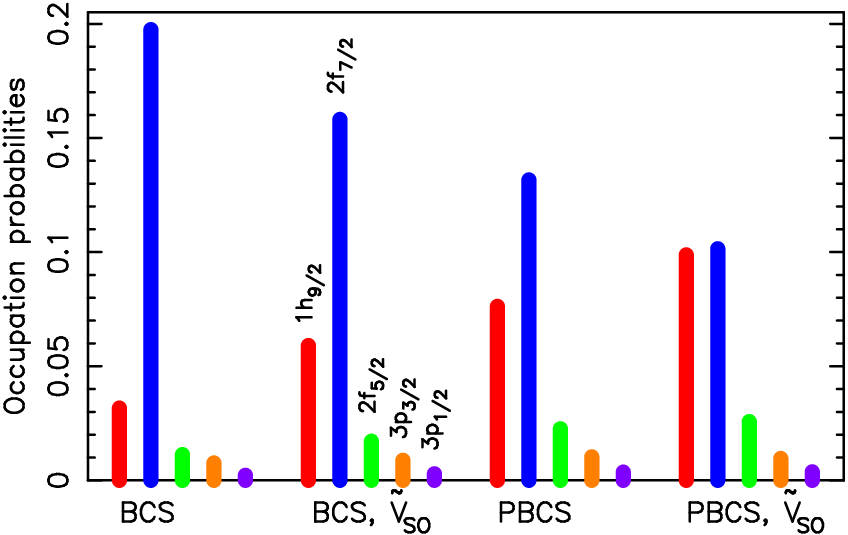}
\caption{\small Occupation probabilities of orbitals above the $N=82$ shell gap in $^{134}$Sn for the four cases indicated in the figure.}
\label{OccuProb}
\end{center}
\end{figure}

The occupation probabilities for the states above the neutron number N=82, which are important for the kink formation, are shown in Fig.~\ref{OccuProb} for $^{134}$Sn. In this nucleus, the pairing calculation involves two neutrons occupying states above those of $^{132}$Sn. In this situation, PBCS provides the exact solution, whereas BCS represents a poor approximation. As shown in Fig.~\ref{OccuProb}, the BCS approach predicts occupation probabilities that differ significantly from the exact PBCS results, most notably overestimating the occupation  of the $2f_{7/2}$ state and underestimating that of $1h_{9/2}$.  This fact has significant consequences for the charge radius
of $^{134}$Sn, as seen in Fig.~\ref{difRc}. Namely, the PBCS predicts a value closer to the experimental value compared to BCS.
 
%Fig. 4

Figure~\ref{difRc} shows that, while the experimental charge radii display a pronounced arch-like 
behavior for $A < 132$, the theoretical calculations fail to reproduce this trend\footnote{We have checked that, within the RMF or relativistic Hartree--Fock approximations with density-independent coupling constants, the charge radii of Sn isotopes lighter than $^{132}$Sn cannot be satisfactorily reproduced. For the relativistic Hartree--Fock approximation, we have explicitly verified this result using the model described in Ref.~\cite{Mar14}, where the pion tensor contribution is reduced to one third of its nominal value, i.e., $\eta' = 1/3$, which we consider to be the most favorable choice for partially accounting for short-range nucleon--nucleon correlation effects.
This behavior of the Sn charge radii appears to be a rather general problem, affecting not only these relativistic models but also the majority of nonrelativistic mean-field models~\cite{Nai23}.}. Unlike the case
of nuclei heavier than $^{132}$Sn, the PBCS and BCS approaches provide similar results for nuclei lighter than $^{132}$Sn. 
This is because, in these nuclei, pairing correlations are predominantly governed by the 
$h_{11/2}$ level. Therefore, in the limit where pairing is restricted to the  
$h_{11/2}$ level, the PBCS and BCS yield identical occupation probability. 

Figure~\ref{difRc} also presents the charge radii calculated using the modified spin--orbit potential, ${\tilde V}^r_{\rm SO}(r,\e)$, Eq.~(\ref{VSOM}). For $A > 132$ these results show a slightly improved agreement with the experimental data compared with those obtained using the original potential, $V^r_{\rm SO}(r,\e)$. 
As will be disccused in more detail below, this improvement arises because the modified potential enhances the occupation of the $1h_{9/2}$ and $1f_{5/2}$, mainly at the expense of the $2f_{7/2}$ orbital. For $^{134}$Sn this enhancement is clearly seen in Fig.~\ref{OccuProb}

To facilitate a quantitative comparison of the kink magnitude across the cases considered in this work, we adopt the quantities defined below \cite{Nai23, Yue24}.

First, we define the charge radius difference
\be
\Delta R_c^Z(A)=R_c^Z(A)-R_c^Z(A-2), 
\ee
where $A$ is the mass number of a nucleus with a magic number of neutrons.
With this quantity, we define the kink indicator \cite{Nai23}
\be
\Delta^2 R_c^Z(A)=\Delta R_c^Z(A+2)-\Delta R_c^Z(A).
\ee
Another kink indicator used in Ref. \cite{Yue24} is defined as
\be
\xi^Z(A)=\frac{\Delta R_c^{Z}(A+2)}{\Delta R_c^{Z}(A)}.
\label{kind}
\ee
Since this work deals exclusively with Sn isotopes, with $Z=50$, we omit below the superscript $Z$
in the quantities defined above.

%Table I
\begin{table} [ht]
\label{xi}
\caption{Values of $\Delta R_c(132)$, $\Delta R_c(134)$, $\Delta^2 R_c$ (in fm), and $\xi$ (dimensionless) 
for the scenarios shown in Fig.~\ref{difRc}.} 
\vspace{0.2 true cm}
%\begin{tabular}{@{\extracolsep{+0.0 mm}} lllll}
%\begin{tabular}{@{\extracolsep{+0.0 mm}} |l|c|c|c|c|}
\begin{tabular}{@{\extracolsep{+0.0 mm}} lcccc}
\hline
 
${\rm case}$ & $\Delta R_c(132) $ & $\Delta R_c(134)$ & $ \Delta^2R_c(132) $ & $ \xi(132)$ \\
\hline
 $ {\rm BCS} $ & $0.0130$ & $0.0152$ & $0.0022$ & $1.17$ \\ 
 $ {\rm BCS,{\tilde V}_{\rm SO}}$  & $0.0125$ & $0.0181$ & $0.0056$ & $1.45$ \\
 $ {\rm PBCS}$ & $0.0128$ & $0.0196$ & $0.0068$ & $1.53$ \\ 
 $ {\rm PBCS,{\tilde V}_{\rm SO}}$ & $0.0122$ & $0.0220$ & $0.0098$ & $1.80$ \\
 $ {\rm Experiment}$ & $0.0071$ & $0.0240$ & $0.0169$ & $3.38$ \\
 \hline
\end{tabular}
\end{table}

As shown in Table I, the discrepancies between the experimental and theoretical values for 
$\Delta R_c(132)$ as well as for the kink indicators (25) and (26) are very large for all the calculations. 
For $\Delta R_c(134)$ the PBCS result obtained with the modified SO potential is in fair
agreement with the experimental value. However, owing to the large discrepancies in  $\Delta R_c(132)$,
even in this case, the kink indicators are poorly reproduced.

The results shown in Table I and Fig.~\ref{difRc} indicate that the main reason for the small kink indicators
in the Sn isotopes is the inability of the RMF model to accurately reproduce the charge radii for $A < 132$. This contrasts with the case of Pb isotopes, for which the model reproduces the charge radii reasonably well for $A < 208$~\cite{Mar24b}.

It is instructive to compare the results presented in Table I with those reported in Table III of Ref.~\cite{Nai23}, 
which correspond to several relativistic Hartree and Hartree-Fock energy density functionals (EDFs) with density-dependent couplings. Modifying the SO interaction in the BCS and PBCS calculations brings our results closer to those obtained within the relativistic Hartree-Fock framework based on the PKO1 and PKO3 parameter sets. 
This behavior suggests that the density dependence introduced in the modified SO potential $\tilde{V}{\rm SO}(r, \e)$, defined in Eq.~(\ref{VSOM}), partially simulates the effect of the density-dependent coupling strength $f_{\pi}$ employed in the PKO1 and PKO3 sets$^1$).
%\cite{note1}.

\begin{figure}[ht]
%\vskip 10 cm
%\leftskip -10 true cm
\begin{center}
%\centering
\includegraphics[width=7.0 cm,angle=0.0]{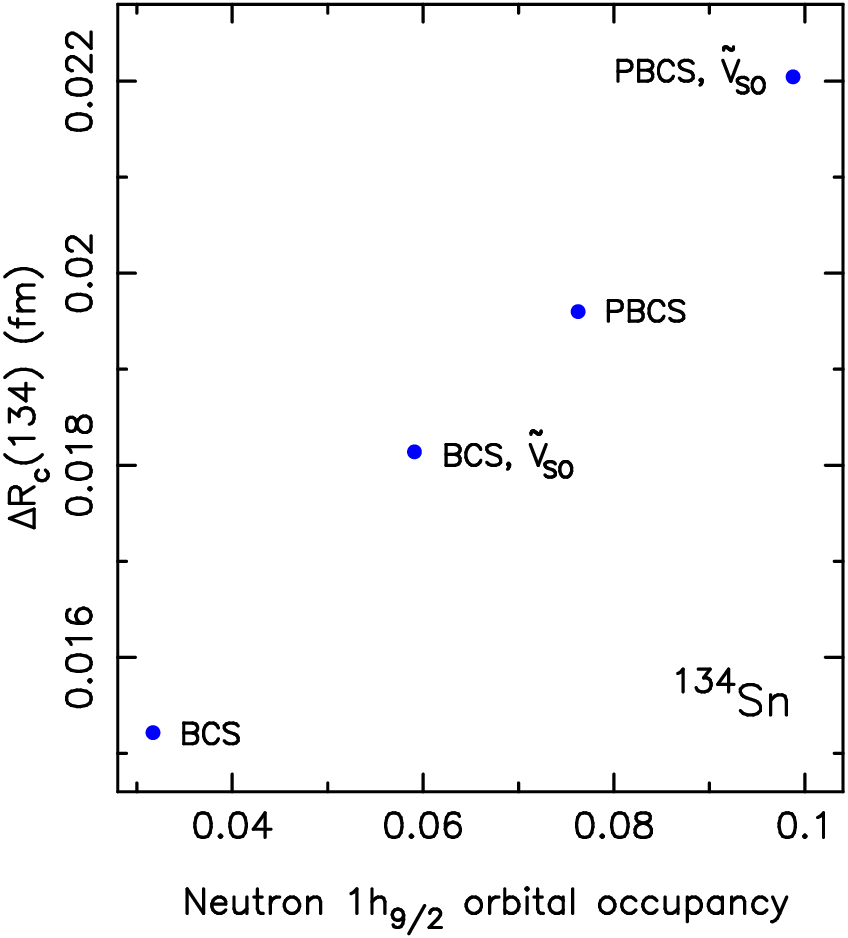}
\caption{\small Values of $\Delta R_c(134)$ as a function of 
the occupation probability of the neutron $1h_{9/2}$ orbital in $^{134}$Sn.}
\label{Ocupro1h92}
\end{center}
\end{figure}

Our results, together with those of other authors \cite{Per21}, show that there is a correlation between the magnitude of the kink in tin isotopes and the occupancy probability of the $1h_{9/2}$ orbital for $A > 132$. 
This correlation can be seen in Fig.~\ref{Ocupro1h92}, which displays $\Delta R_c(134)$ as a function of the occupation
probability of the $1h_{9/2}$ orbital in $^{134}$Sn for the four cases analyzed in this work.

%Fig. 5

\subsection{Charge radii with different neutron configurations}

\begin{figure}[ht]
%\vskip 10 cm
%\leftskip -10 true cm
\begin{center}
%\centering
\includegraphics[width=8.2 cm,angle=0]{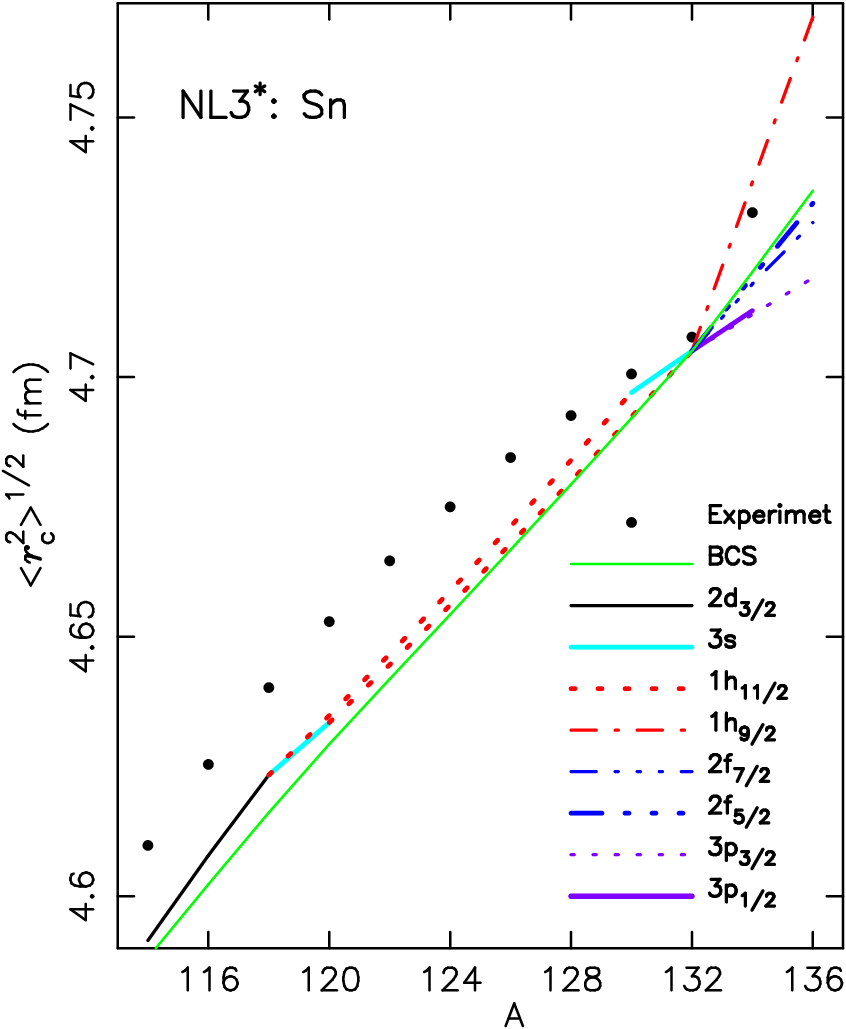}
\caption{\small The rms charge radii $R_c\equiv \langle r_c^2 \rangle^{1/2}$ of $^{A}$Sn isotopes, as in Fig.~\ref{difRc}, 
%for several neutron configurations in the interval $114 \leq A \leq 132$ (the $2d1h3s$-conf. and $2d3s1h$-conf.) and $A>132$ (see the text).
for the neutron $2d1h3s$ and $2d3s1h$ configurations in the interval $114 \leq A \leq 132$, and the last five configurations listed at the bottom of the figure ($1h_{9/2}-3p_{1/2}$) for $A>132$ (see text).
Note that, for $118 < A < 132$, $R_c (2d3s1h) < R_c(2d1h3s)$.}
\label{Rc}
\end{center}
\end{figure}

The next question we address is whether alternative neutron configurations, different from those that emerge from 
the RMFA energy spectrum, could lead to improved predictions for the charge radii. 
The considered neutron configurations (without pairing) and the corresponding charge radii, calculated self-consistently 
using the NL3$^*$ parameter set, are shown in Fig.~\ref{Rc}.
%Fig. 6

For the $^A$Sn isotopes with $114 < A \le 132$, the predicted order of neutron level occupancy in the RMFA is $2d_{3/2}$, $3s_{1/2}$, and $1h_{11/2}$.
We refer to this specific filling sequence as the $2d3s1h$ configuration ($2d3s1h$-conf.).

As shown in Fig.~\ref{Rc}, the results obtained with this configuration fail to reproduce the charge radii.
A modest improvement is observed  with the filling sequence $2d_{3/2}$, $1h_{11/2}$, and $3s_{1/2}$, referred
to as the $2d1h3s$ configuration ($2d1h3s$-conf.).

The behavior of the charge radii for $A > 132$ depends strongly on the configuration of the valence neutrons.
Figure~\ref{Rc} shows the results for the charge radii assuming that the valence neutrons occupy one of the following
orbitals: $1h_{9/2}$, $2f_{7/2,\;5/2}$, or $3p_{3/2,\;1/2}$.
It can be observed that neutrons in the orbital $1h_{9/2}$ ($1h_{j_-}$-conf.) lead to a much larger charge radius compared to those in the orbitals $2f_{j_+}$ ($2f_{j_+}$-conf.) and $2f_{j_-}$ ($2f_{j_-}$-conf.). 
Moreover, the increase in the charge radii due to neutrons in these orbitals is considerably larger than that resulting from neutrons in the 
$3p_{3/2}$ and $3p_{1/2}$ orbitals ($3p_{j_+}$-conf. and $3p_{j_-}$-conf., respectively).
The reasons for this behavior will be discussed below (see Subsec. III. F). Our results for the $1h_{j_-}$ and $2f_{j_+}$ configurations agree with those reported in Fig. 13 of Ref. \cite{Per21}.

The conclusions of this analysis are the following: 
(i) there is no neutron configuration that, within the RMFA, 
provides a reasonable description of the arch-like behavior of the charge radii for Sn isotopes with $A < 132$; 
(ii) for $A > 132$, the charge radii depend strongly on the occupancy of the $1h_{9/2}$ orbital, as also pointed out in Ref. \cite{Per21}. 

The second conclusion (ii) suggests that the kink in the charge radii could be enhanced by reducing the SO potential in the low-density surface region, where the $1h_j$ orbitals still have significant amplitudes (see Fig.~\ref{G2F2} below), thereby leaving the more deeply bound orbitals largely unaffected. As shown in Fig.~\ref{VSOVSOM}, our choice for $\tilde V_{\rm SO}$ in Eq.~(\ref{VSOM}) satisfies this requirement by reducing the magnitude of $\tilde V_{\rm SO} \times r^2$ relative to that of $V_{\rm SO} \times r^2$ for $r \gtrsim 5$~fm.

\subsection{Charge radii of proton orbitals for different neutron configurations}

\subsubsection{Nodeless proton orbitals}

\begin{figure}[ht]
%\vskip 10 cm
%\leftskip -10 true cm
\begin{center}
%\centering
\includegraphics[width=9. cm,angle=270]{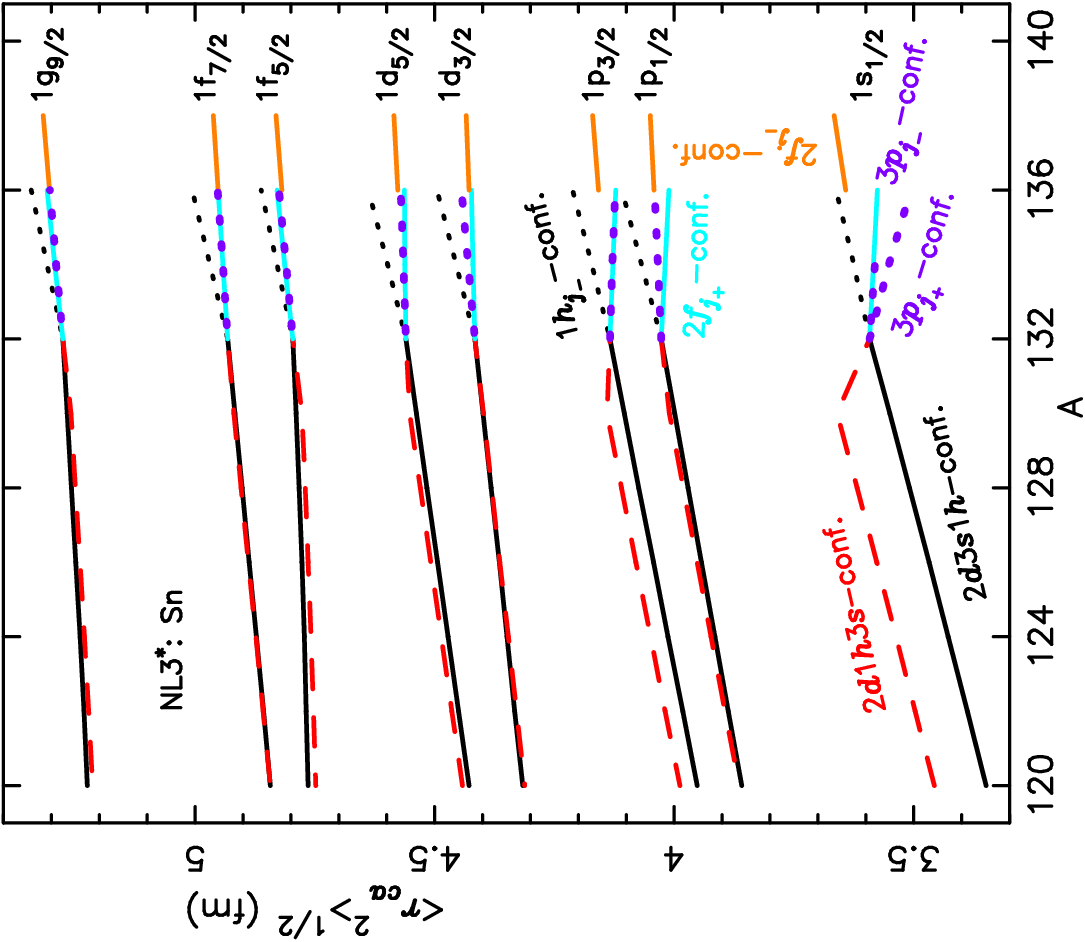} 
\caption{\small The rms charge radii, $\langle {r_{ca}}^2\rangle^{1/2}$, of the nodeless ($n=1$) proton orbitals $a$, labeled as $nl_j$ on the right-hand side of the figure, are plotted as a function of $A$.
For $A<132$, neutrons are assumed to occupy the $3s_{1/2}$, $2d_{3/2}$, and $1h_{11/2}$ orbitals, following the occupation sequences of the two configurations considered, labeled as $2d1h3s$-conf. and $2d3s1h$-conf. 
For $A > 132$, neutrons are assumed to occupy one of the following orbitals: $1h_{9/2}$ (denoted as the $1h_{j_-}$-conf.), $2f_{7/2}$ ($2f_{j_+}$-conf.), $2f_{5/2}$ ($2f_{j_-}$-conf.), $3p_{3/2}$ ($3p_{j_+}$-conf.), or $3p_{1/2}$ ($3p_{j_-}$-conf.).
The $3p_{j_-}$-conf. is considered only for the $1s_{1/2}$ proton orbital. For proton orbitals with $l \ge 1$, the results are similar to those of the $3p_{j_+}$-conf. and are therefore not shown.
For the $2f_{j_-}$-conf., only the results for $^{136}$Sn and $^{138}$Sn are displayed for clarity. The values for $^{134}$Sn can be inferred by extrapolation.}
\label{rci0}
\end{center}
\end{figure}

To better understand the evolution of charge radii with atomic mass number $A$, we now examine the variation of the charge radii of individual proton sp states for different neutron configurations.
We begin with the charge radii of the nodeless sp proton states shown in Fig. \ref{rci0}.

%Fig. 7

For $A < 132$, the figure indicates that the charge radii of proton states with small angular momentum ($l \leq 2$) vary significantly with the neutron configuration. 

For $A > 132$, the charge radii of proton states are considerably larger in the $1h_{j_-}$-conf. than in the others. 
The effects of valence neutrons occupying the $2f_{j_+}$ and $2f_{j_-}$ orbitals on the charge radii of sp states differ significantly only for proton states with angular momentum $l \leq 1$. 

\subsubsection{Proton orbitals with nodes}

\begin{figure}[ht]
%\vskip 10 cm
%\leftskip -10 true cm
\begin{center}
%\centering
\includegraphics[width=9. cm,angle=270]{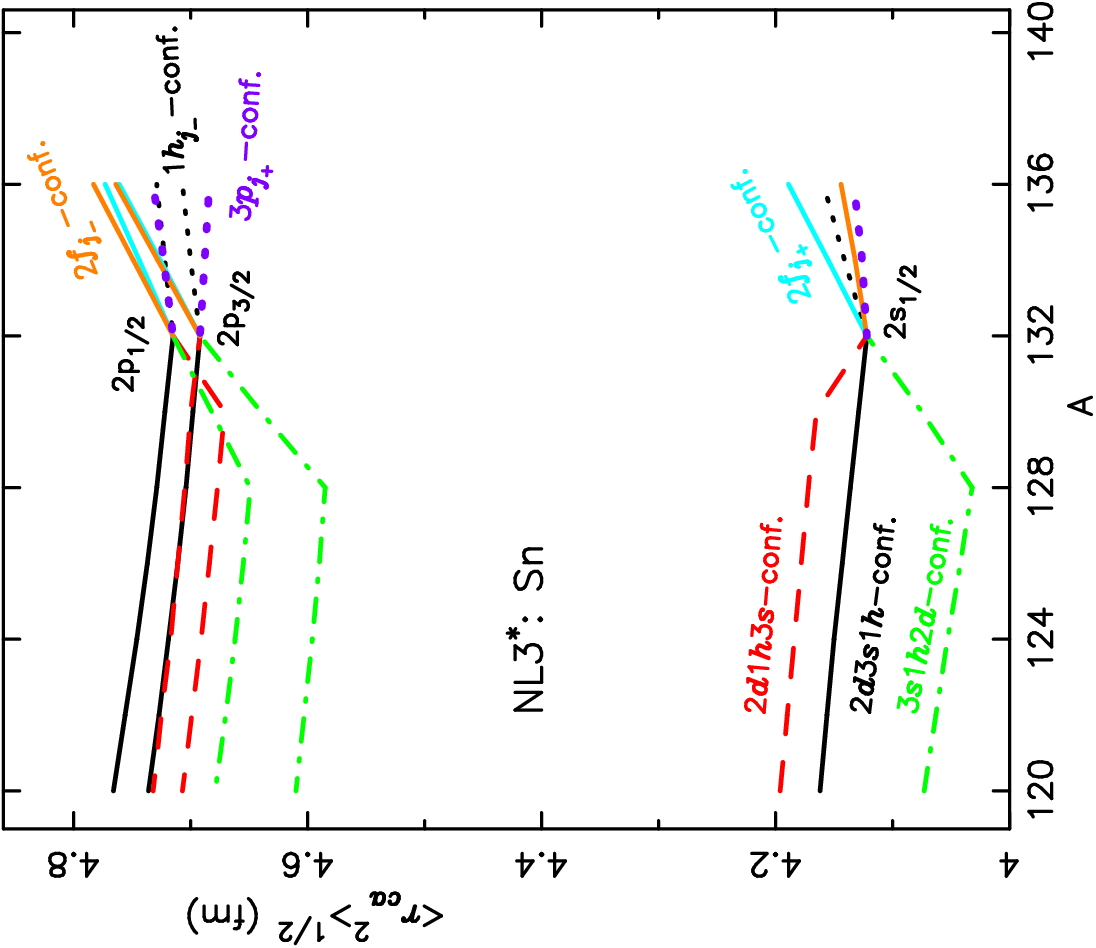} 
\caption{\small Same as Fig.~\ref{rci0} (including also the results for the $3s1h2d$-conf.) but for proton orbital with nodes. 
Note that for the $2p$ proton orbitals, the contribution to $\langle {r_{ca}}^2\rangle^{1/2}$ from neutrons in the $2f_{j_+}$-conf. are always smaller than those in the $2f_{j_-}$-conf.
The results for the $3p_{j_-}$-conf. are nearly identical to those of the $3p_{j_+}$-conf. and are therefore omitted.}
\label{rcin}
\end{center}
\end{figure}

We now discuss the charge radii of the nodal proton orbitals $2s_{1/2}$ and $2p_{1/2,\;3/2}$. 
Their dependence on various neutron configurations is shown in Fig.~\ref{rcin}. 

%Fig. 8

A comparison of Figs.~\ref{rci0} and \ref{rcin} reveals that, unlike in the case of Pb isotopes \cite{Nie12, God13}, the contribution of nodal orbitals to the kink in Sn isotopes remains significant.
In fact, to understand the relatively steep slope of $\langle r_c^2 \rangle^{1/2}$ in Fig.~\ref{Rc} when the valence neutrons fill the 
$2d_{3/2}$ orbital, it is necessary to consider the contribution of the eight protons in the $2s_{1/2}$ and $2p_{1/2,\;3/2}$ orbitals.

\subsection{Effect of the Dirac spinor components of valence neutrons on the nuclear charge radii}

In this subsection, we analyze the effect of the small components of the Dirac spinors corresponding to the valence neutrons on
the charge radii in the kink region. 

To focus the discussion on the relevant aspects, we adopt a simplified assumption: for $122 \leq A \leq 132$, all the valence neutrons are assumed to occupy the $1h_{11/2}$ orbital, whereas for isotopes with $A > 132$, the additional neutrons occupy only one of the following orbitals: $1h_{9/2}$, $2f_{7/2}$, $2f_{5/2}$, or $3p_{3/2}$.

\begin{figure}[ht]
%\vskip 10 cm
%\leftskip -10 true cm
\begin{center}
%\centering
\includegraphics[width=8.0 cm,angle=0]{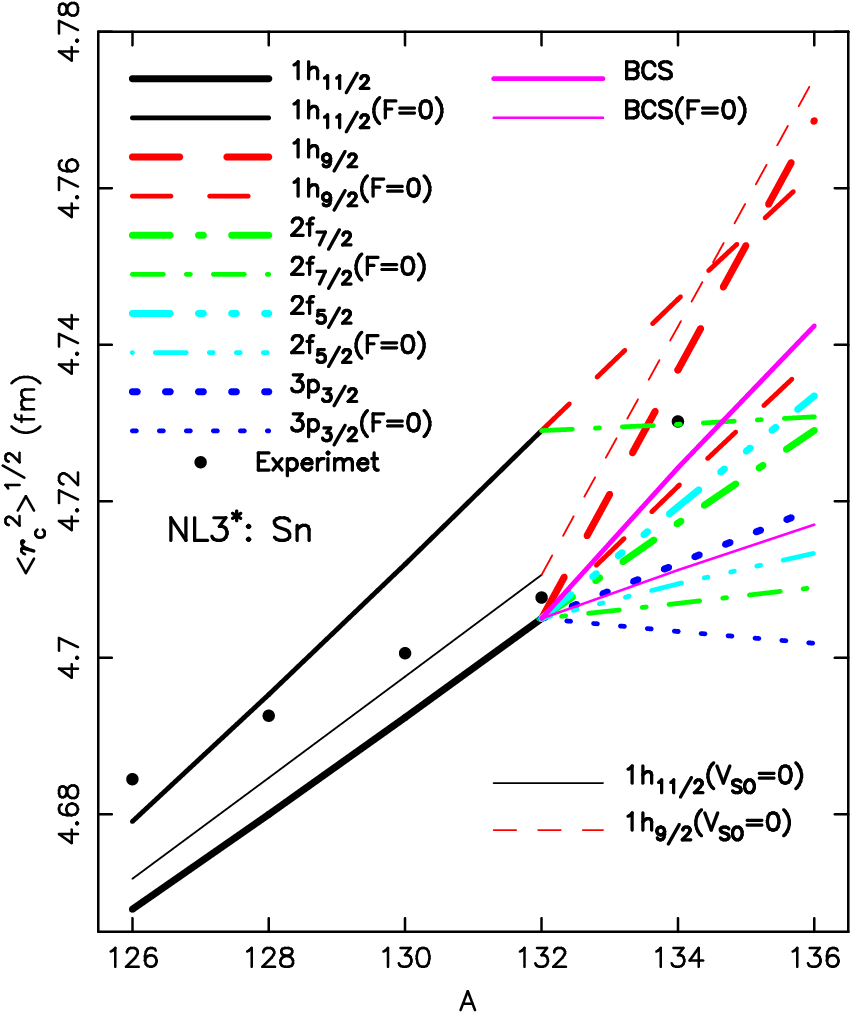} 
\caption{\small Charge radii $R_c \equiv \langle r_c^2 \rangle^{1/2}$ for the valence neutrons occupying the orbitals indicated in the figure. 
Results labeled ($F=0$) correspond to calculations in which the small component is set to zero, 
while labels $1h_j(V_{\rm SO}=0)$ denote calculations with the spin--orbit interaction neglected for the $1h$ orbitals.}
\label{RcF}
\end{center}
\end{figure}

The RMF results for the charge radii obtained for these configurations are shown in Fig.~\ref{RcF}. 
The figure also includes the radii calculated when the contribution of the small components $F_a$ is neglected throughout the self-consistent procedure. In this case, the normalization condition (\ref{nor}) is imposed on the large component $G_a$. 

Since the radial structures of the large and small components of the  Dirac spinor differ substantially, the variation of $G_a$ arising from its modified normalization when $F_a$ is neglected cannot compensate for the effects of $F_a$ on the nucleus, particularly on the charge radius. More importantly, as discussed in the next paragraph, even if the contribution of $F_a$ to the density were exactly compensated
by a corresponding change in the normalization of $G_a$, the omission of $F_a$ from the single-particle central potential cannot be compensated by such a modification of $G_a$. This is mainly, though not exclusively, because their respective contributions to the central potential enter with opposite signs.

%Fig. 9

Fig.~\ref{RcF} shows that, for $A \leq 132$, the contribution of the small component $F_{1h_{11/2}}$ to $R_c$ is significant, 
slightly improving the agreement {\it with the}  experimental charge radii.
For $A > 132$, it can be observed that the contributions of the small components $F_{1h_{9/2}}$, $F_{2f_{{7/2},\;{5/2}}}$, and $F_{3p_{3/2}}$ to $R_c$ are also important.
In fact, since for $A > 132$ the occupancies of both orbitals $1h_{9/2}$ and $2f_{7/2}$ are significant when pairing correlations are considered, the components $F_{1h_{9/2}}$ and $F_{2f_{7/2}}$ play a crucial role in determining the magnitude of the kink.

To better appreciate the contributions of various neutron configurations on the charge radii and the 
role played by the small components $F_a$, in Table~II are shown the values 
of $\Delta R_c(134)$, $\Delta^2R_c(132)$, and $\xi$  for each case.
They provide a quantitative measure of the contribution of neutrons, {\it when assumed to occupy specific orbitals}, to the kink in the charge radius.
Larger values of these quantities indicate a more pronounced kink.
%[-The rows labeled ``Differences'' show the contributions of the small components of the valence neutron orbitals to the quantities indicated at the top of each column.-]
As expected from Fig.~\ref{Rc}, for a given number of neutrons, the largest contributions to $\Delta R_c(134)$, $\Delta^2 R_c(132)$, and $\xi$ come from neutrons occupying the $1h_{9/2}$ orbital. 
Some of these contributions will be discussed in detail below.

%Table II
\begin{table} [ht]
\label{xi2}
\caption{
Values of $\Delta R_c(134)$, $\Delta^2 R_c(132)$ (in fm), and $\xi(132)$ calculated with valence neutrons occupying the orbitals indicated in the first column, either with their small components $F \neq 0$ or $F = 0$. 
Below these results, their differences are also shown. 
In the fourth row of each orbital's section, we display the results obtained with the SO interaction neglected ($V_{\mathrm{SO}} = 0$) and $F \neq 0$. 
In all cases, the value of $\Delta R_c(132)$ is 0.01265 fm for $V_{\mathrm{SO}} \neq 0$ and 0.0130 fm for $V_{\mathrm{SO}} = 0$. Both results are obtained under the assumption that, for $120 < A \leq132$, neutrons occupy the $1h_{11/2}$ orbital. Including BCS pairing yields very similar values (see Table I).}

%\begin{tabular}{@{\extracolsep{+0.0 mm}} lllll}
\begin{tabular}{@{\extracolsep{1.0 mm}} lccc}
\hline
 $\nu$-orbital  & $\Delta R_c(134)$ & $ \Delta^2R_c(132) $ & $ \xi(132)$ \\
\hline
 $ 1h_{9/2},\; F\ne0 $ & 0.0318 & 0.0191 & 2.511 \\
 \hskip 1.0 cm $ F=0 $ & 0.0169 & 0.0042 & 1.335 \\ 
 $\;\;\;\;\;\;\;\;\;$ Differences & 0.0149 & 0.0149 & 1.176 \\
 $\;\;\;\;\;\;\;\;\;\;V_{\rm SO}=0, F\neq0$& 0.0317 & 0.0187 & 2.442 \\
\hline
 $ 2f_{5/2},\; F\ne0 $ & 0.0143 & 0.0016 & 1.128 \\ 
 \hskip 0.9 cm $ F=0 $ & 0.0044 &-0.0082 & 0.350 \\ 
 $\;\;\;\;\;\;\;\;\;$ Differences           & 0.0098 & 0.0098 & 0.778 \\ 
 $\;\;\;\;\;\;\;\;\;\;V_{\rm SO}=0, F\neq0$& 0.0140 & 0.0013 & 1.107 \\ 
 \hline
 $ 2f_{7/2},\; F\ne0 $ & 0.0122 &-0.0005 & 0.962 \\ 
 \hskip 0.9 cm $ F=0 $ & 0.0019 &-0.0107 & 0.153 \\ 
 $\;\;\;\;\;\;\;\;\;$ Differences & 0.0102 & 0.0102 & 0.809 \\ 
 $\;\;\;\;\;\;\;\;\;\;V_{\rm SO}=0, F\neq0$& 0.0128 & 0.00012 & 1.012 \\ 
 \hline
 $ 3p_{1/2},\; F\ne0 $ & 0.0078 &-0.0049 & 0.616 \\ 
 \hskip 0.9 cm $ F=0 $ &-0.0014 &-0.0141 &-0.111 \\ 
$\;\;\;\;\;\;\;\;\;$ Differences & 0.0092 & 0.0092 & 0.727 \\ 
 $\;\;\;\;\;\;\;\;\;\;V_{\rm SO}=0, F\neq0$& 0.00805 &-0.0046 & 0.6364 \\ 
 \hline
 $ 3p_{3/2},\; F\ne0 $ & 0.0070 &-0.0056 & 0.555 \\ 
 \hskip 0.9 cm $ F=0 $ &-0.0016 &-0.0143 &-0.130 \\ 
 $\;\;\;\;\;\;\;\;\;$ Differences & 0.0087 & 0.0087 & 0.685 \\ 
 $\;\;\;\;\;\;\;\;\;\;V_{\rm SO}=0, F\neq0$& 0.00686 & -0.0058 & 0.5423 \\ 
 \hline
 
\end{tabular}
\end{table}

\begin{figure}[ht]
%\leftskip -10 true cm
\begin{center}
%\centering
\includegraphics[width=5.5 cm,angle=-90.0]{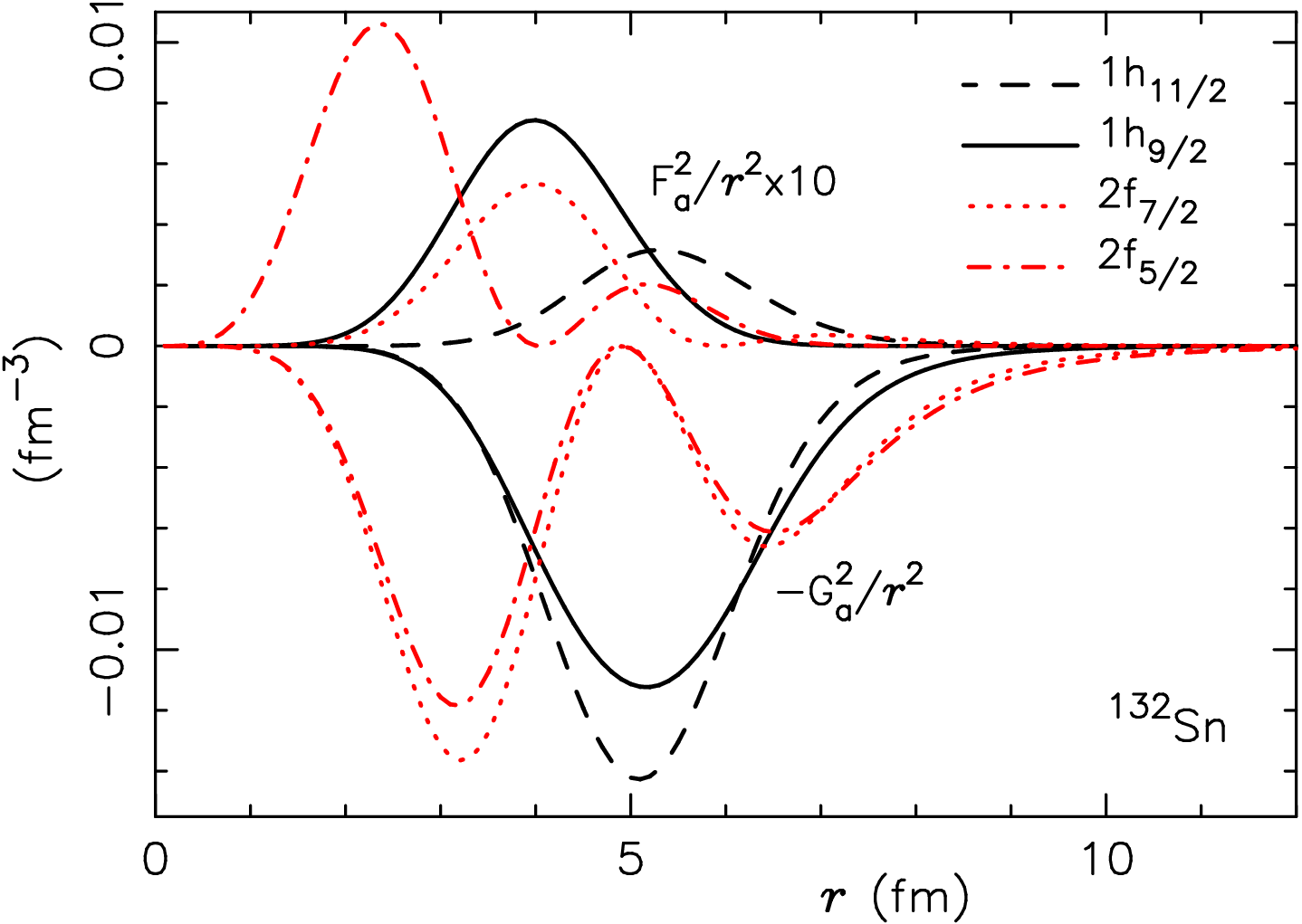}
\caption{\small 
Functions $-G_a^2/r^2$ (negative) and $F_a^2/r^2$ (positive) corresponding to the orbitals indicated in the figure for $^{132}$Sn. 
}
\label{G2F2}
\end{center}
\end{figure}

The behavior of the charge radii around the kink region shown in Fig.~\ref{difRc} is mainly related to the localisation
properties of the $1h_{11/2,\;9/2}$ and $2f_{7/2,\;5/2}$ neutron orbits. 
Figure~\ref{G2F2} shows the square of the radial parts of the small and large components of the Dirac spinors for these neutron orbitals, calculated for $^{132}$Sn.
For convenience,  
the large components are plotted with the opposite sign\footnote{It can be seen that, in the local density approximation, the contribution to the sp central potential $V_{\rm cent}$ in the Schr\"odinger equation (\ref{Sch}) of the large and small components
of a sp spinor are proportional to $-G_a^2/r$ and $F_a^2/r$, respectively, see Eq. (\ref{Vna2}). }.  
We note that the large components of the SO doublets $1h_{11/2,9/2}$ and $2f_{7/2,5/2}$ are quite similar, as expected, whereas their small components differ considerably.
Figure~\ref{G2F2} also shows that, although the large components of the $1h_{9/2}$ and $2f_{7/2}$ orbitals 
are very different, their small components are rather similar. 
This feature is characteristic of pseudospin doublets~\cite{Gin98, Mar05, Mar08}.
Note that, in the absence of the SO interaction, the radial parts of the large components of two SO partner states are proportional to each other (see Sec.~II). Therefore, the differences displayed in Fig.~\ref{G2F2} between the radial parts of the large components of the $1h$ and $2f$ SO 
partners arise primarily from the SO interaction.

%Fig. 10

The charge-radius slopes in Fig.~\ref{RcF} around $A = 132$, for neutrons in the $1h$-conf. and $2f$-conf., 
are governed by the geometrical characteristics of the $1h$ and $2f$ orbitals.
In particular, the slope of the charge radii calculated without the small components is almost insensitive to whether neutrons occupy the $1h_{11/2}$ or $1h_{9/2}$ orbitals, reflecting the similarity of their large components $G_{1h_{11/2}}$ and $G_{1h_{9/2}}$.

In contrast, the notable change in slope between $A= 132$ and $A = 134$ observed 
in the full calculations is related to the differences between the 
small components of the SO doublet\footnote{When a kink in an isotopic family is primarily caused by the filling of two levels belonging to the same spin--orbit doublet with neutrons, the previous result helps us to understand why producing a sufficiently large kink is more challenging with nonrelativistic models than with relativistic ones. This is roughly the case for the Sn and Pb isotopic chains, where the kinks appear at $N = 82$ and $N = 126$, respectively.}. 
In fact, if we neglect the SO interaction for all $1h$ states while 
retaining their small components, we obtain the two lines labeled 
$1h_{11/2}(V_{\rm SO}=0)$ and $1h_{9/2}(V_{\rm SO}=0)$.  
We observe that they are almost parallel to the corresponding result for $V_{\rm SO} \neq 0$ (note that the results for $V_{\rm SO} = 0$ and $V_{\rm SO} \neq 0$ must converge at $A = 120$).
This behaviour is consistent with the results for the orbital $1h_{9/2}$ shown in Table II. More precisely,
it can be seen that the quantities $\Delta R_c(134)$, $\Delta^2 R_c(132)$,  and $\xi(132)$ do not change 
much when the SO interaction is put to zero. 

\begin{figure}[ht]
%\leftskip -10 true cm
\begin{center}
%\centering
\includegraphics[width=5.5 cm, angle=-90.0]{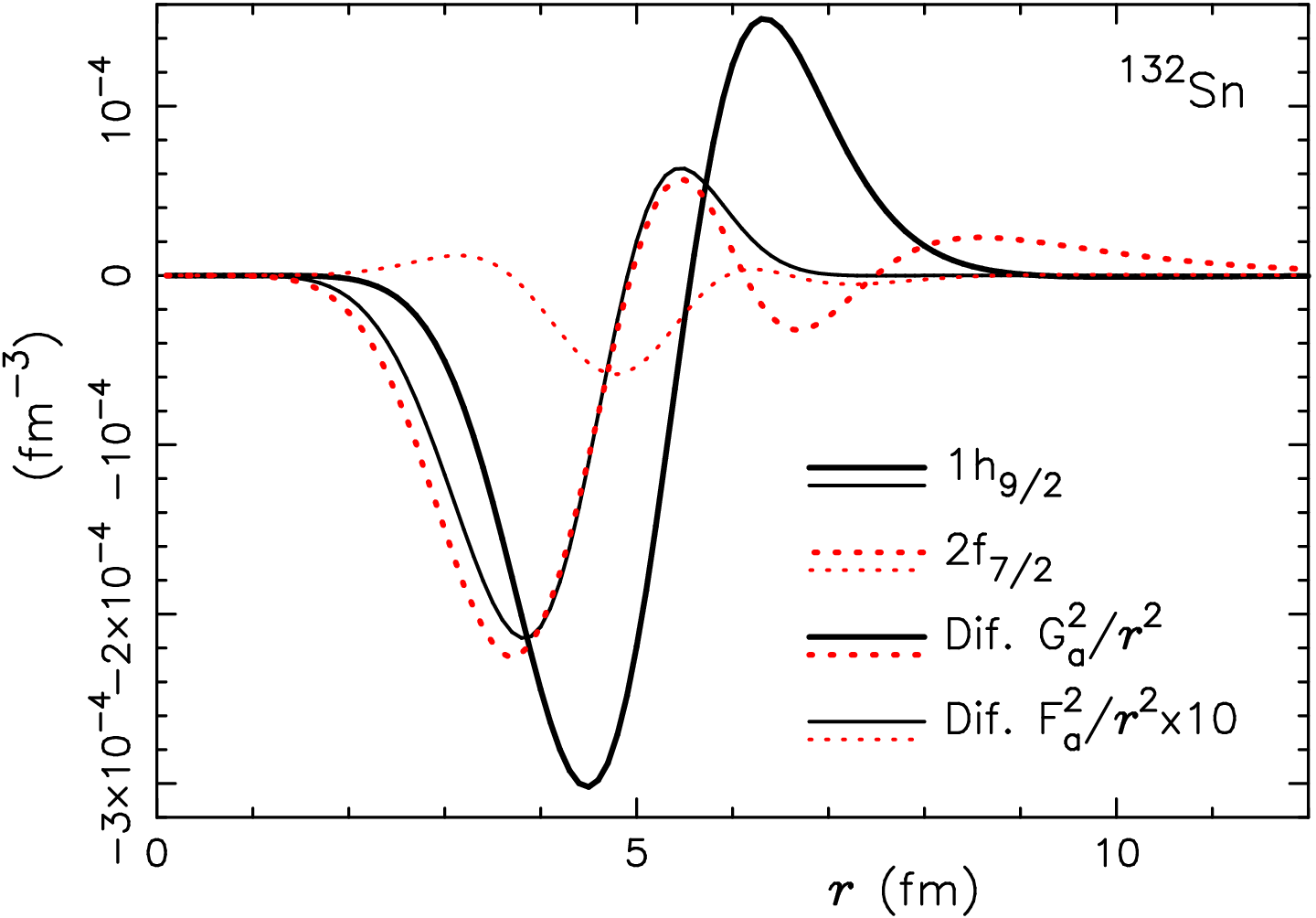}
\caption{\small Differences of $G_a^2/r^2$ calculated with the modified SO potential (Eq.~(\ref{VSOM})) and the unmodified potential (Eq.~(\ref{VSO})) for the orbitals $a=1h_{9/2}$ and $a=2f_{7/2}$. The same is shown for $F_a^2/r^2$.}
\label{DifG2F2}
\end{center}
\end{figure}

We have checked that the same conclusions hold for the modified SO interaction $\tilde V_{\rm SO}$. 
This is because  by replacing $V_{\rm SO}$ with $\tilde V_{\rm SO}$ the wave functions of the valence neutron orbitals do not change significantly (see Fig.~\ref{DifG2F2}). Due to this reason, the improved
kink description obtained by replacing $V_{\rm SO}$ with $\tilde V_{\rm SO}$ is not related to the
changes in the wave functions but with the increased occupation probability of the orbital $1h_{9/2}$.

The results discussed above indicate why the traditional nonrelativistic mean-field  
models have difficulty describing the kink arising from the occupancy of the $1h_{11/2}$ 
and $1h_{9/2}$ orbitals. Since the $1h_{9/2}$ orbital plays an important role in the kink 
formation, the SO interaction must be weak enough to allow a sizable occupancy 
of this orbital. On the other hand, the SO interaction should be strong enough to induce
significant differences between the wave functions of the $1h_{11/2}$ and $1h_{9/2}$ orbitals.
These two conditions are difficult to fulfil simultaneously in the non-relativistic models. 
In the case of RMF the second condition is not necessary because the contribution
of the small components to the charge radii, important for the kink emergence, is not
depending much on the strength of the SO interaction.

%Fig. 11

\subsection{Contribution of components of the valence neutron spinors to the proton central potential}

To understand how the components of the occupied neutron spinors affect the charge radii, we have analysed their contributions to the proton central potential $V_{\rm cent}(r, \e)$ defined in Eq. (\ref{Vcent}). 
This potential includes the state-dependent relativistic term, $V^{\rm rel}_{\rm cent}(r, \e)$, which is proportional to $\frac{1}{M}$.  
Since this term is relatively small and varies from one proton state to another, is not considered 
in the present analysis, as it would unnecessarily complicate the discussion.
Therefore, we focus in the following on the main part of the proton central potential
$V_{\rm cent^*}(r) \equiv S(r) + V(r)$ (see Eqs.~(\ref{Sch})--(\ref{Vcentrel})). 
In the local density approximation, before self-consistency is achieved, the contribution of one neutron in the $a$ state to this potential can be approximated by \cite{Mar24b}:
\bea
\begin{split}
V^{{\rm n},a}_{\rm cent^*}(r)\simeq &\left[\frac{g_\omega^2}{m_\omega^2}-\frac{g_\rho^2}{m_\rho^2}-\frac{g_\sigma^2}{{m_\sigma^*}^2}\right]\frac{G_a^2}{4\pi r^2}\\
+&\left[\frac{g_\omega^2}{m_\omega^2}-\frac{g_\rho^2}{m_\rho^2}+\frac{g_\sigma^2}{{m_\sigma^*}^2}\right]\frac{F_a^2}{4\pi r^2}.
\end{split}
\label{Vna2}
\eea

For the NL3$^*$ parameter set \cite{Mar24b}, the factor multiplying the small component of the spinor is approximately 
five times larger than that for the large component (a similar ratio is found with other parameter sets within the RMFA). 
This fact significantly enhances the contribution of the small components to the central potential, which otherwise would
be negligible compared with that of the large component. This behavior contrasts with that of the baryon density, where the
factors multiplying both components are identical (see Eq.~(\ref{eqdn})).

The difference between the factors multiplying the large and small components in 
Eq.~(\ref{Vna2}) arises from the fact that the scalar potential 
$S \simeq - g_{\sigma}^2 / m_{\sigma}^{*2}\, \rho_S < 0$ 
and the vector potential 
$V \simeq g_{\omega}^2 / m_{\omega}^2\, \rho_B - g_{\rho}^2 / m_{\rho}^2\, \rho_{0,3} > 0$ 
are both very large in magnitude compared with their difference~\cite{Mar24b}; that is, 
$|S| \sim V \gg |S| - V $ or, equivalently, $V - S \gg V + S$ (see Eqs. (\ref{Dir})-(\ref{eqd3}), (\ref{Sch})-(\ref{Vcentrel})). 
In symmetric nuclear matter at the equilibrium density for the NL3$^*$ parameter set, 
$S \simeq -380~\text{MeV}$, $V \simeq 310~\text{MeV}$, and $V_{cent}\simeq S + V \simeq -70~\text{MeV}$.

At this point it is important to mention that the small components of the nucleon Dirac spinors do not play a significant role in the SO potential \cite{Mar24b}.
 
Below, we examine how the presence of valence neutrons in different orbitals modifies the self-consistent proton central potential $V_{\rm cent^*}(r)$ and, in turn, how these modifications affect the nuclear charge radius. 
As an illustrative example, we consider the effect of adding four neutrons to $^{132}$Sn in various neutron orbitals.

\subsubsection{Effect of neutron $1h$ orbitals on the charge radius}

We begin with the case in which four neutrons are added to the orbital $1h_{9/2}$. 
The variation of the proton central potential induced by these four neutrons, relative  to $^{132}$Sn, in which the orbit $h_{9/2}$ is empty, is given by
\bea
\delta V_{\rm cent^*}(r)=\frac{4}{A_1-A_2} \bigg[[V_{\rm cent^*}(r)]_{^{A_1}\rm Sn} \nonumber \\
-[V_{\rm cent^*}(r)]_{^{A_2}\rm Sn}\bigg],
\label{dVce}
\eea
with $A_1=136$ and $A_2=132$. 

\begin{figure}[ht]
%\vskip 10 cm
%\leftskip -10 true cm
\begin{center}
%\centering
\includegraphics[width=5.0 cm,angle=270]{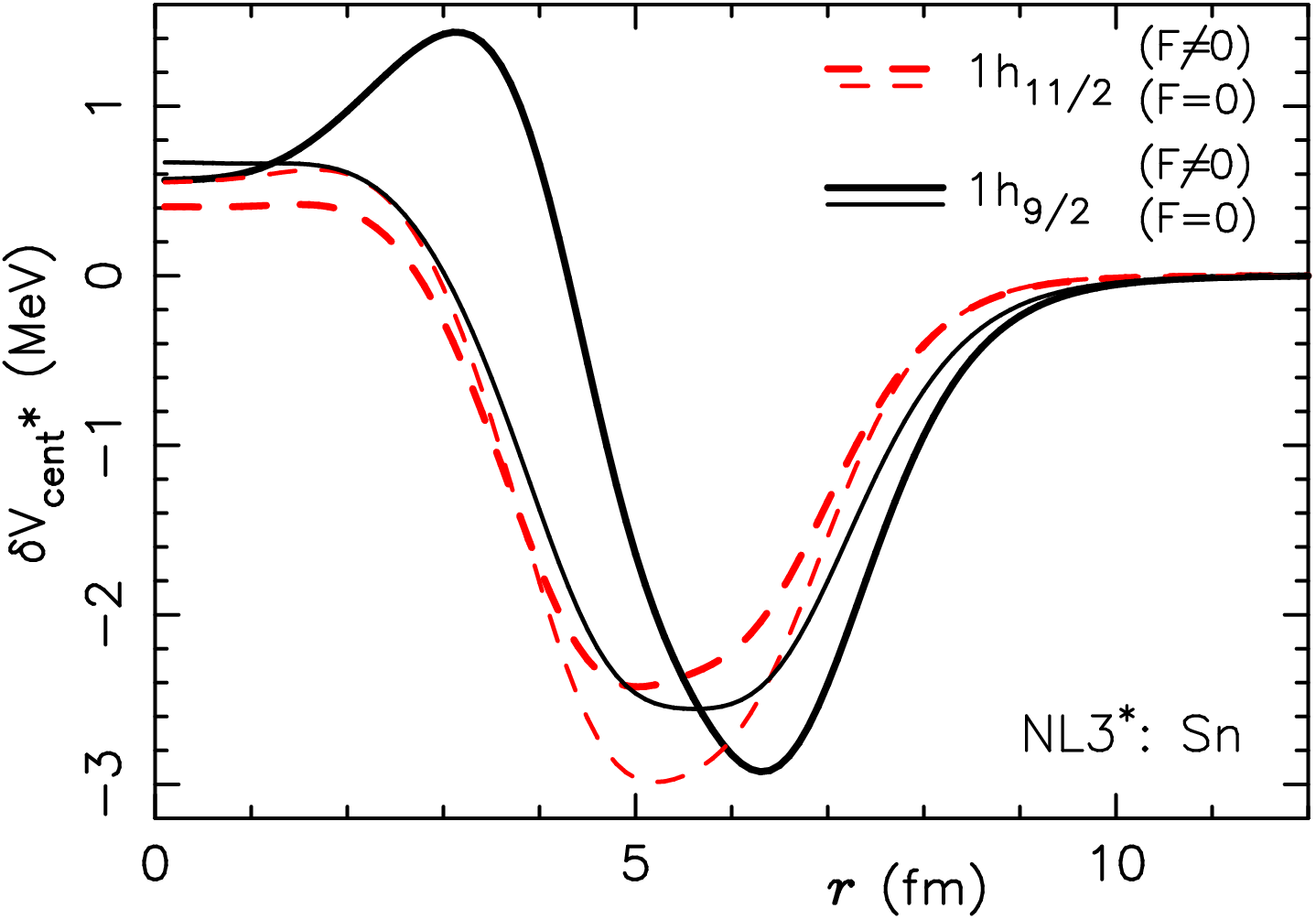} 
\caption{\small
Variation of the proton central potential, $\delta V_{\rm cent^*}(r)$,
induced by four neutrons occupying alternatively the $1h_{11/2}$ orbital of $^{132}$Sn
and the $1h_{9/2}$ orbital of $^{136}$Sn.
Note that $\delta V_{\rm cent^*}(r)$ is identical for all proton orbitals.
}
\label{dVcene1h}
\end{center}
\end{figure}

The variation of the proton potential calculated with $F_{1h_{9/2}}(r)\ne 0$ and $F_{1h_{9/2}}(r)=0$ is shown in Fig.~\ref{dVcene1h}. 
The inclusion of $F_{1h_{9/2}}(r)$ produces two main effects: 
(i) a significant increase of the potential within the nuclear interior, generating a peak around $r = 3$~fm; and 
(ii) a reduction of the potential near the surface, accompanied by an outward shift of its minimum. 
These combined effects displace protons toward the nuclear surface, leading to a pronounced increase in the charge radius when $F_{1h_{9/2}}(r) \ne 0$, 
as shown in Fig.~\ref{RcF}. 
Furthermore, this figure indicates that the magnitude of the effect grows approximately with $(A-132)$, i.e., it is roughly proportional to the number of neutrons added to the $1h_{9/2}$ orbital.
%Fig. 12

For comparison, Fig.~\ref{dVcene1h} also illustrates how the proton central potential changes due to the presence of four neutrons in the orbital $h_{11/2}$.
%(which represents one-third of the twelve neutrons occupying the $h_{11/2}$ orbital in $^{132}$Sn ).
 The variation of the proton potential due to the four neutrons is estimated relative to $^{120}$Sn, in which the orbit $h_{11/2}$ is empty in the absence of pairing. The results, calculated with Eq. (\ref{dVce}) for $A_1=132$ and $A_2=120$, are shown in Fig.~\ref{dVcene1h}. 
It can be seen that the contribution of the small component $F_{1h_{11/2}}$ slightly reduces the potential in the inner part of the nucleus and increases it significantly near the minimum. 
These two effects of $F_{1h_{11/2}}$ drive the protons toward the interior of the nucleus, resulting in a decrease of the charge radii, as shown in Fig.~\ref{RcF}. 
%From this figure, it can be noticed that this effect on the radii decreases as $A$ varies from 120 to 132. 
Note that the two lines in this figure corresponding to the $1h_{11/2}$ orbital (with $F_{1h_{11/2}} = 0$ and $F_{1h_{11/2}} \neq 0$) coincide at $A = 120$, 
since the $1h_{11/2}$ orbital begins to be occupied for $A > 120$, and their separation thereafter increases approximately proportionally to $(A-120)$.

\subsubsection{Effect of neutron $2f$ orbitals on the charge radius}

\begin{figure}[ht]
%\vskip 10 cm
%\leftskip -10 true cm
\begin{center}
%\centering
\includegraphics[width=5.0 cm,angle=270]{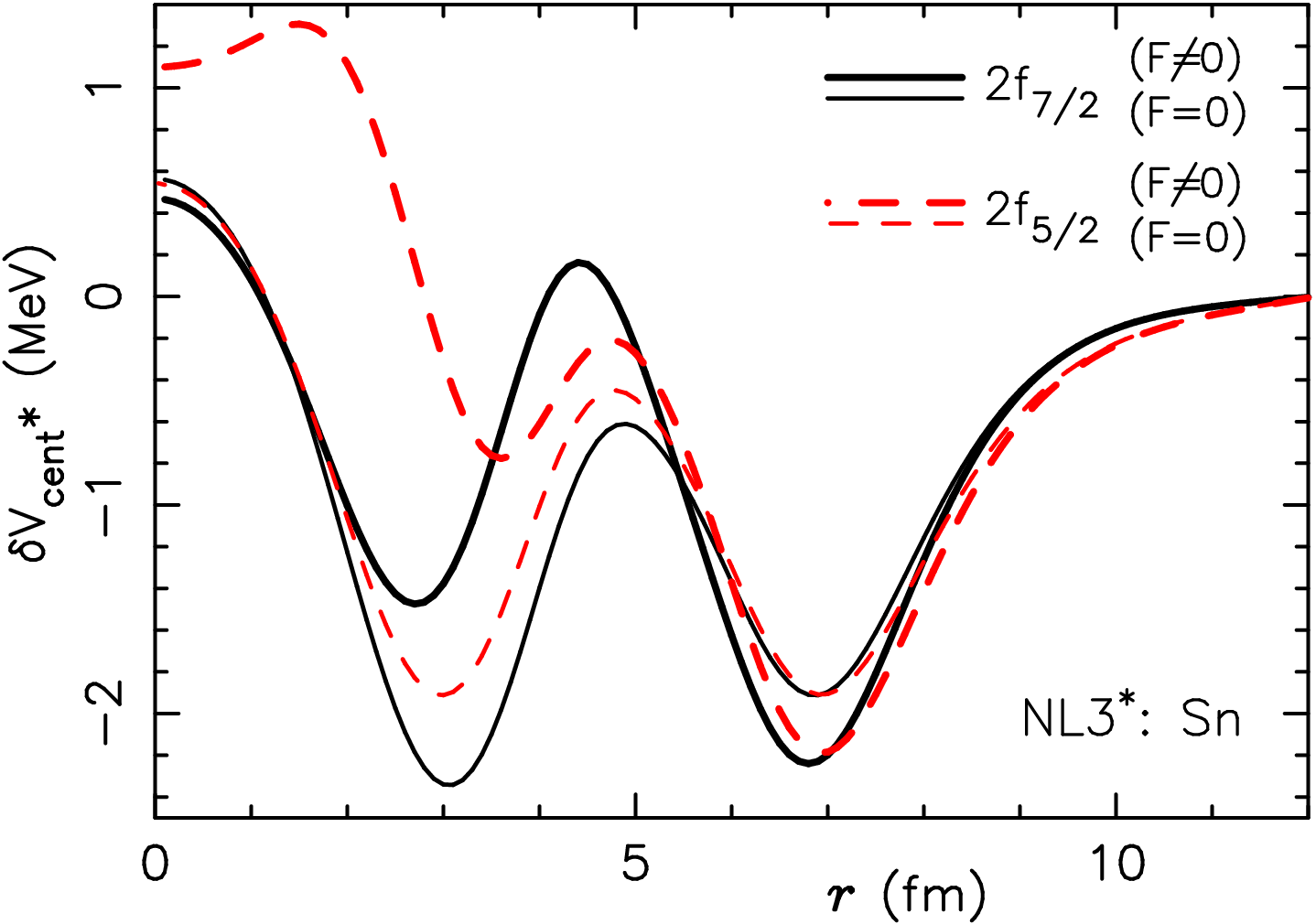} 
\caption{\small Same as Fig.~\ref{dVcene1h} but for the neutron $2f_{5/2,\;7/2}$ orbitals in $^{136}$Sn.}
\label{dVcene2f}
\end{center}
\end{figure}

Figure~\ref{dVcene2f} shows the contribution to $V_{\rm cent^*}(r)$  from four neutrons in $^{136}$Sn occupying either the $2f_{7/2}$ or $2f_{5/2}$ orbital.
It can be seen that the small component $F_{2f_{7/2}}(r)$ increases the proton central potential in the region $2 \lesssim r \lesssim 5$~fm, 
thereby making the first minimum shallower, and decreases it around the second minimum, making it deeper. 
This contribution clearly leads to an increase in the charge radii, as shown in Fig.~\ref{RcF}.

%Fig. 13

In the case of the neutrons filling the $2f_{5/2}$ orbital, the 
small component $F_{2f_{5/2}}(r)$ enhances the potential in the inner region of the
nucleus, while leaving it nearly unchanged for $r \gtrsim 5$~fm, where the second minimum of the potential is located. As a result, $F_{2f_{5/2}}(r)$ also contributes to the increase of the
charge radii, as illustrated in Fig.~\ref{RcF}. 

Figure~\ref{dVcene2f} shows that the small-component contributions to the proton central potential differ markedly between the $2f_{7/2}$ and $2f_{5/2}$ orbitals, with the latter tending to favor larger radii.
Nevertheless, the overall impact of these two orbitals on the nuclear radius is not significantly different, since the region within the nucleus where their contributions diverge is relatively limited.

\subsubsection{Effect of neutron $3p$ orbitals on the charge radius}

\begin{figure}[ht]
%\vskip 10 cm
%\leftskip -10 true cm
\begin{center}
%\centering
\includegraphics[width=5.0 cm,angle=270]{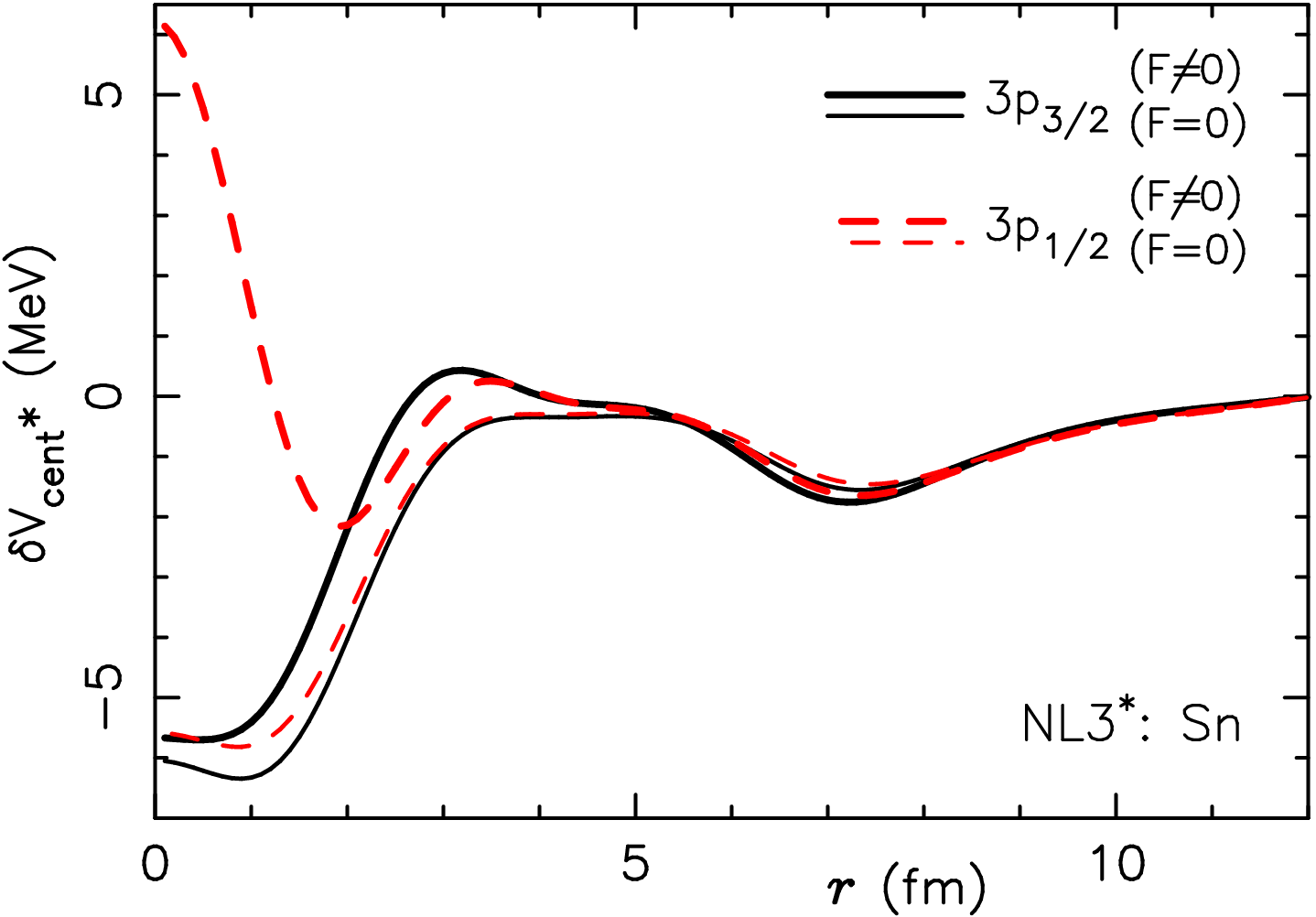} 
\caption{\small
Variation of the proton central potential, $\delta V_{\rm cent^*}(r)$, 
induced alternatively by four neutrons occupying the $3p_{3/2}$ orbital of $^{136}$Sn 
and by two neutrons occupying the $3p_{1/2}$ orbital of $^{134}$Sn.
The latter result is multiplied by two to allow a direct comparison between the two orbitals.
}
\label{dVcene3p}
\end{center}
\end{figure}

Figure~\ref{dVcene3p} shows the variation of $V_{\rm cent^*}(r)$ as neutrons occupy the $3p_{3/2}$ and $3p_{1/2}$ orbitals. 
These results indicate that valence neutrons in the $3p_{1/2}$ orbital enhance $V_{\rm cent^*}(r)$ in the inner region of the nucleus much more than those in the $3p_{3/2}$ orbital, due to the contribution from the small component of their Dirac spinor. 
Nevertheless, this enhancement is confined to a limited region within the nuclear interior.
Consequently, the overall impact on the charge radii is small, as can be seen in Fig.~\ref{Rc}. 

%Fig. 14

The results from Figs.~\ref{RcF}, \ref{dVcene1h}, \ref{dVcene2f}, and \ref{dVcene3p} show that the increase in charge radii is more pronounced when neutrons occupy the SO orbitals with $j = j_-$ than when they occupy the corresponding SO partners with $j = j_+$.
This behavior arises primarily from the greater magnitude of the small component of the Dirac spinor in the inner
region of the nucleus for orbitals with $j = j_-$, as shown in Fig.~\ref{G2F2}.
This is because in Eq. (\ref{F}) the quantum number $k$ is positive for orbitals 
with $j = j_-$ and negative for those with $j = j_+$.
As a result, for neutron orbitals with $j = j_-$, the two terms in the numerator interfere constructively at small values of $r$, whereas for those with $j = j_+$, they interfere destructively in this region. 
The influence of neutrons in these two types of orbitals on a given proton orbital depends on the spatial distribution of the latter. 
In most cases, the effect is more pronounced for proton orbitals with high probability density in the inner regions of the nucleus, where the small components of neutron spinors reach their maximum values.

The pronounced difference in the spatial structure of the small components of SO partner orbitals in the inner region of the nucleus is one of the  key features of relativistic models responsible for the emergence of the kink effect, although these models do not always fully reproduce its magnitude.

Another key feature is the specific structure of the sp central potential in relativistic models.
This potential contains a strong attractive scalar term, originating from the exchange of effective $\sigma$ mesons between nucleons, and a strong repulsive vector term, arising mainly from the exchange of $\omega$ and $\rho$ mesons (with the Coulomb contribution being comparatively small).
The coexistence of these two large terms with opposite signs markedly enhances the impact of the small components of the neutron spinors on the proton potential.

The small components of the Dirac spinor are not only relevant for the specific structure of the single-particle SO orbitals and of the central potential, but are in fact deeply connected with the very mechanism by which relativistic models achieve nuclear saturation. In these frameworks, they play a fundamental role in the saturation mechanism itself.
By contrast, in nonrelativistic approaches saturation is obtained through the density dependence of the energy density functional (EDF), so that the effects associated with the small components are effectively absorbed into the parametrization of the functional. 

Because saturation properties directly influence nuclear sizes, the evolution of the charge radius with increasing mass number strongly depends on the underlying saturation mechanism of the model.
One may therefore ask whether the mechanisms responsible for saturation in relativistic and nonrelativistic frameworks are approximately equivalent in determining the evolution of charge radii with atomic mass. As we have demonstrated in this work, they are far from equivalent. The difference becomes particularly evident when two SO partner orbitals play a dominant role in the formation of the kink, as observed in the tin isotopes. If one retains only the contribution of these two orbitals and neglects the SO interaction, relativistic models can still produce a kink owing to the presence of the small components of the Dirac spinors, whereas standard nonrelativistic models cannot.

\section{Summary and Conclusions}

In this work, we analyzed the evolution of the root-mean-square charge radii ($R_c$) along the Sn isotopes
using the relativistic mean-field approximation with the NL3$^*$ parameter 
set.

Discrepancies are observed in the evolution of $R_c$  between the experimental data and the theoretical 
predictions for $ A<132 $. While the experimental charge radii display a pronounced arch-like behavior in this region, the theoretical values fail to reproduce this feature. We have explored alternative neutron configurations obtained by reordering the single-particle levels; however, none of them leads to a significant improvement in the agreement with the data.

The RMFA is likewise unable to accurately reproduce the magnitude of the kink in the charge radii around $^{132}$Sn. 
For nuclei heavier than  $^{132}$Sn, the charge radii are reasonably well described when a slightly weakened spin--orbit (SO) interaction at the nuclear surface is adopted, and pairing is treated within the PBCS approximation. Even in this case, however, the magnitude of the kink remains underestimated. This shortcoming originates from the inadequate description of the charge radii for nuclei lighter than $^{132}$Sn.

In the present study, particular attention was given to the relativistic effects associated 
with the small components of the Dirac spinors. 
It was shown that the small components of the orbitals corresponding to the valence neutrons  drive the emergence of the kink through their contributions to the proton central potential.
 In particular, due to differences in the small components of the SO orbitals, the kink increases
more strongly when the valence neutrons fill the SO orbitals with  $j = j_-$ than when they fill the corresponding SO partners with $j = j_+$.
However, this increase is not sufficient, by itself, to fully account for the experimentally observed magnitude of the kink. In fact, the discrepancy mainly arises from the underestimated charge radii 
of tin isotopes with $A < 132$.

In conclusion, although the standard RMF fails to accurately reproduce the charge radii of Sn isotopes, 
it does predict the emergency of a kink. In RMF this kink arises as a genuine relativistic 
effect stemming from the small components of the Dirac spinors corresponding to the valence neutrons.

As far as we know, the only relativistic model which can provide accurate results for 
Sn isotopes is the RHF with the density-dependent coupling constants \cite{Wan21, Nai23}. 
In these calculations, it appears that the tensor force, generated by the Fock term, plays an important
role in the accurate prediction of the charge radii and of the kink behavior. On the other hand, we have
checked that the RHFA performed with the standard density-independent coupling constant provide
results similar to those of the RMFA presented in this study. It could be interesting to investigate in details
the reasons why the two RHF approached provide so different results for the charge radii in Sn isotopes. 
This is a study beyond the scope of this paper.

\end{document}